\documentclass[12pt,nohyper]{JHEP3}
\usepackage{graphics,psfrag}
\usepackage{amsmath,amsthm,amssymb,epsfig,euscript,array,cite}
\usepackage{cases,empheq,pmat}
\usepackage{eufrak}

\newtheorem{thm}{Theorem}[section]
\newtheorem{lem}[thm]{Lemma}

\theoremstyle{definition}

\theoremstyle{remark}

\newcounter{multieqs}

\newcommand{\be}{\begin{equation}}
\newcommand{\ee}{\end{equation}}
\newcommand{\eq}[1]{(\ref{#1})}
\newcommand{\bit}{\begin{itemize}}  \newcommand{\eit}{\end{itemize}}

\newcommand{\bm}[1]{\mbox{\boldmath $#1$}}
\newcommand{\rf}[1]{(\ref{#1})}

\def\bd{\begin{document}}
\def\ed{\end{document}}
\def\nn{\nonumber}
\def\bea{\begin{eqnarray}}
\def\eea{\end{eqnarray}}
\let\bm=\bibitem

\def\la{\langle}
\def\ra{\rangle}

\def\npb#1#2#3{Nucl. Phys. {\bf{B#1}} #3 (#2)}
\def\plb#1#2#3{Phys. Lett. {\bf{#1B}} #3 (#2)}
\def\prl#1#2#3{Phys. Rev. Lett. {\bf{#1}} #3 (#2)}
\def\prd#1#2#3{Phys. Rev. {D \bf{#1}} #3 (#2)}
\def\cmp#1#2#3{Comm. Math. Phys. {\bf{#1}} #3 (#2)}
\def\cqg#1#2#3{Class. Quantum Grav. {\bf{#1}} #3 (#2)}
\def\nppsa#1#2#3{Nucl. Phys. B (Proc. Suppl.) {\bf{#1A}}#3 (#2)}
\def\ap#1#2#3{Ann. of Phys. {\bf{#1}} #3 (#2)}
\def\ijmp#1#2#3{Int. J. Mod. Phys. {\bf{A#1}} #3 (#2)}
\def\rmp#1#2#3{Rev. Mod. Phys. {\bf{#1}} #3 (#2)}
\def\mpla#1#2#3{Mod. Phys. Lett. {\bf A#1} #3 (#2)}
\def\jhep#1#2#3{J. High Energy Phys. {\bf #1} #3 (#2)}
\def\atmp#1#2#3{Adv. Theor. Math. Phys. {\bf #1} #3 (#2)}

\def\N{{\cal N}}
\def\sst{\scriptscriptstyle}
\def\thetabar{\bar\theta}
\def\Tr{{\rm Tr}}
\def\one{\mbox{1 \kern-.59em {\rm l}}}

\def\a{\alpha}      \def\da{{\dot\alpha}}  \def\dA{{\dot A}}
\def\b{\beta}       \def\db{{\dot\beta}}  
\def\g{\gamma}  \def\G{\Gamma}  \def\dc{{\dot\gamma}}  
\def\d{\delta}  \def\D{\Delta}  \def\ddt{\dot\delta}  
\def\e{\epsilon}        \def\ve{\varepsilon}  
\def\f{\phi}    \def\F{\Phi}    \def\vvf{\f}  
\def\h{\eta}  
\def\k{\kappa}  
\def\l{\lambda} \def\L{\Lambda}  
\def\m{\mu} \def\n{\nu}  
\def\o{\omega}  
\def\p{\pi} \def\P{\Pi}  
\def\r{\rho}  
\def\s{\sigma}  \def\S{\Sigma}  
\def\t{\tau}  
\def\th{\theta} \def\Th{\Theta} \def\vth{\vartheta}  
\def\X{\Xeta}  
\def\z{\zeta}  

\def\na{\nabla}  

\def\cA{{\cal A}} \def\cB{{\cal B}} \def\cC{{\cal C}}  
\def\cD{{\cal D}} \def\cE{{\cal E}} \def\cF{{\cal F}}  
\def\cG{{\cal G}} \def\cH{{\cal H}} \def\cI{{\cal I}}  
\def\cJ{{\cal J}} \def\cK{{\cal K}} \def\cL{{\cal L}}  
\def\cM{{\cal M}} \def\cN{{\cal N}} \def\cO{{\cal O}}  
\def\cP{{\cal P}} \def\cQ{{\cal Q}} \def\cR{{\cal R}}  
\def\cS{{\cal S}} \def\cT{{\cal T}} \def\cU{{\cal U}}  
\def\cV{{\cal V}} \def\cW{{\cal W}} \def\cX{{\cal X}}  
\def\cY{{\cal Y}} \def\cZ{{\cal Z}}

\def\ua{\underline{\alpha}}  
\def\uc{\underline{\phantom{\alpha}}\!\!\!\gamma}  
\def\um{\underline{\mu}}  
\def\ud{\underline\delta}  
\def\ue{\underline\epsilon}  
\def\una{\underline a}\def\unA{\underline A}  
\def\unb{\underline b}\def\unB{\underline B}  
\def\unc{\underline c}\def\unC{\underline C}  
\def\und{\underline d}\def\unD{\underline D}  
\def\une{\underline e}\def\unE{\underline E}  
\def\unf{\underline{\phantom{e}}\!\!\!\! f}\def\unF{\underline F}  
\def\unm{\underline m}\def\unM{\underline M}  
\def\unn{\underline n}\def\unN{\underline N}  
\def\unp{\underline{\phantom{a}}\!\!\! p}\def\unP{\underline P}  
\def\unq{\underline{\phantom{a}}\!\!\! q}  
\def\unQ{\underline{\phantom{A}}\!\!\!\! Q}  
\def\unH{\underline{H}}  
  
\def\As {{A \hspace{-6.4pt} \slash}\;}  
\def\bs {{b \hspace{-6.4pt} \slash}\;}  
\def\Ds {{D \hspace{-6.4pt} \slash}\;}  
\def\ds {{\del \hspace{-6.4pt} \slash}\;}  
\def\ss {{\s \hspace{-6.4pt} \slash}\;}  
\def\ks {{ k \hspace{-6.4pt} \slash}\;}  
\def\ps {{p \hspace{-6.4pt} \slash}\;}   
\def\xs {{x \hspace{-6.4pt} \slash}\;}  
\def\pas {{{p_1} \hspace{-6.4pt} \slash}\;}  
\def\pbs {{{p_2} \hspace{-6.4pt} \slash}\;}   
\def\cFs {{{\cal F} \hspace{-6.4pt} \slash}\;}

\def\Dh{{\hat{D}}}
\def\Gh{{\hat{G}}}
\def\Fh{{\hat{F}}}
\def\Ih{{\hat{I}}} 
\def\Jh{{\hat{J}}} 
\def\Kh{{\hat{K}}}
\def\Lh{{\hat{L}}} 
\def\Ph{{\hat{P}}}
\def\Rh{{\hat{R}}}
\def\Vh{{\hat{V}}} 
\def\Xh{{\hat{X}}}
 
\def\ah{{\hat{\a}}}
\def\bh{{\hat{\b}}}
\def\gh{{\hat{\g}}}
\def\dh{{\hat{\d}}}
\def\hh{\hat{h}}
\def\uh{\hat{u}}  
\def\xh{\hat{x}}  
\def\yh{\hat{y}}  
\def\ph{\hat{p}}  
\def\xih{\hat{\xi}}  
\def\chih{\hat{\chi}}

\def\psit{\tilde{\psi}}  
\def\Psit{\tilde{\Psi}}   
\def\Psibt{\tilde{\bar{Psi}}}  

\def\st{\tilde{\sigma}}  

\def\Phit{\tilde{\Phi}}   
\def\Phitb{\overline{\tilde{Phi}}}  
\def\tht{\tilde{\th}}  
\def\lt{\tilde{\l}}
\def\chit{\tilde{\chi}}   
\def\phit{\tilde{\phi}} 

\def\At{\tilde{A}}
\def\Bt{\tilde{B}}
\def\Ct{\tilde{C}}
\def\Dt{\tilde{D}}
\def\Et{\tilde{E}}
\def\Ft{\tilde{F}}
\def\Ht{\tilde{H}}
\def\It{\tilde{I}}
\def\Jt{\tilde{J}}
\def\Qt{\tilde{Q}}  
\def\Rt{\tilde{R}}  
\def\Mt{\tilde{M }}  
\def\Nt{\tilde{N}}   
\def\St{\tilde{S}}
\def\Vt{\tilde{V}}
\def\Xt{\tilde{X}} 
\def\at{\tilde{a}}
\def\ct{\tilde{c}}   
\def\htt{\tilde{h}} 
\def\ft{\tilde{f}}
\def\gt{\tilde{g}}
\def\pt{\tilde{p}}  
\def\qt{\tilde{q}}  
\def\vt{\tilde{v}}  
\def\nt{\tilde{n}}  
\def\ut{\tilde{u}}  
\def\wt{\tilde{w}}  
\def\zt{\tilde{z}} 
\def\xt{\tilde{x}} 
\def\yt{\tilde{y}} 
\def\Psit{\tilde{\Psi}}
\def\vphit{\tilde{\varphi}}  

\def\delb{\bar{\partial}}  
\def\thb{\bar{\theta}}
\def\mub{\bar{\mu}}
\def\lamb{\bar{\l}}
\def\psib{\bar{\psi}}
\def\sb{\bar{\sigma}}
\def\xib{\bar{\xi}}
\def\chib{\bar{\chi}}

\def\Phib{\bar{\Phi}}
\def\Lamb{\bar{\Lambda}}
\def\Sb{{\overline \Sigma}}
\def\cb{\bar{c}}
\def\hb{\bar{h}}
\def\qb{\bar{q}}
\def\wb{\bar{w}}
\def\ub{\bar{u}}
\def\zb{{\bar{z}}}
\def\Hb{\bar{H}}
\def\Qb{{\bar Q}}

\def\Ab{{\overline A}} \def\Bb{{\overline B}} \def\Cb{{\overline C}}  
\def\Db{{\overline D}} \def\Eb{{\overline E}} \def\Fb{{\overline F}}  
\def\Gb{{\overline G}} 
\def\Ib{{\overline I}}  
\def\Jb{{\overline J}} \def\Kb{{\overline K}} \def\Lb{{\overline L}}  
\def\Mb{{\overline M}} \def\Nb{{\overline N}} \def\Ob{{\overline O}}  
\def\Pb{{\overline P}}  \def\Rb{{\overline R}}  
 \def\Tb{{\overline T}} \def\Ub{{\overline U}}  
\def\Vb{{\overline V}} \def\Wb{{\overline W}} \def\Xb{{\overline X}}  
\def\Yb{{\overline Y}} \def\Zb{{\overline Z}}  

\def\fb{{\overline f}}
\def\gb{{\overline g}}
\def\mb{{\overline m}}
\def\lb{{\overline l}}
\def\yb{{\overline y}}

\def\ba{{\bf a}} 
\def\bk{{\bf k}}  
\def\bl{{\bf l}}  
\def\bp{{\bf p}}  
\def\bq{{\bf q}}  
\def\br{{\bf r}}
\def\bt{{\bf t}}
\def\bu{{\bf u}}
\def\bv{{\bf v}}
\def\bx{{\bf x}}  
\def\by{{\bf y}}  
\def\bR{{\bf R}}  
\def\bV{{\bf V}}

\def\bone{{\bf 1}}  

\def\va{{\vec a}}
\def\vk{{\vec k}}
\def\vp{{\vec p}}
\def\vq{{\vec q}}
\def\vx{{\vec x}}
\def\vy{{\vec y}}
\def\vu{{\vec u}}
\def\vv{{\vec v}}

\def\vs{{\vec \sigma}}
\def\vtau{{\vec \tau}}

\newcommand{\ov}[1]{\overrightarrow{#1}}

\def\frA{\mathfrak{A}}
\def\frB{\mathfrak{B}}
\def\frC{\mathfrak{C}}
\def\frD{\mathfrak{D}}
\def\frE{\mathfrak{E}}
\def\frF{\mathfrak{F}}
\def\frG{\mathfrak{G}}
\def\frH{\mathfrak{H}}
\def\frM{\mathfrak{M}}
\def\frN{\mathfrak{N}}
\def\frR{\mathfrak{R}}
\def\frW{\mathfrak{W}}

\def\fra{\mathfrak{a}}
\def\frb{\mathfrak{b}}
\def\frf{\mathfrak{f}}
\def\frg{\mathfrak{g}}
\def\frh{\mathfrak{h}}
\def\frl{\mathfrak{l}}
\def\frs{\mathfrak{s}}
\def\fri{\mathfrak{i}}
\def\frj{\mathfrak{j}}

\def\ma{\mathfrak{a}}
\def\mg{\mathfrak{g}}
\def\mh{\mathfrak{h}}
\def\mR{\mathfrak{R}}
\def\mN{\mathfrak{N}}

\def\d{\delta}\def\D{\Delta}\def\ddt{\dot\delta}  
  
\def\pa{\partial} \def\del{\partial}  
\def\xx{\times}  
\def\uno{\mbox{1 \kern-.59em {\rm l}}}    
  
\def\trp{^{\top}}  
\def\inv{^{-1}}  
\def\dag{{^{\dagger}}}  
\def\pr{^{\prime}}  
  
\def\rar{\rightarrow}  
\def\lar{\leftarrow}  
\def\lrar{\leftrightarrow}  
  
\newcommand{\0}{\,\!}      
\def\one{1\!\!1\,\,}  
\def\im{\imath}  
\def\jm{\jmath}  
  
\newcommand{\tr}{\mbox{tr}}  
\newcommand{\slsh}[1]{/ \!\!\!\! #1}  
  
\def\vac{|0\rangle}  
\def\lvac{\langle 0|}  
  
\def\hlf{\frac{1}{2}}  
\def\ove#1{\frac{1}{#1}}  

\def\Box{\square}  
\def\CC {\mathbb{C}}
\def\FF {\mathbb{F}}
\def\RR{\mathbb{R}}
\def\NN{\mathbb{N}}  
\def\ZZ{\mathbb{Z}}  
\def\bb#1{{\bf #1}}  
\def\bcomment#1{}  
\def\bfhat#1{{\bf \hat{#1}}}  
\def\VEV#1{\left\langle #1\right\rangle}  

\newcommand{\ex}[1]{{\rm e}^{#1}} \def\ii{{\rm i}}  

\newcommand{\lrbrk}[1]{\left(#1\right)}
\newcommand{\sfrac}[2]{{\textstyle\frac{#1}{#2}}}
 
\def\stw{{\sqrt{2}}}

\def\rf {{\rm f}}
\def\ri {{\rm i}}
\def\rj {{\rm j}}
\def\rk {{\rm k}}
\def\rl {{\rm l}}
\def\rs {{\scriptscriptstyle \rm S}}
\def\rt {{\scriptscriptstyle \rm T}}

\def\rQ {{\scriptscriptstyle \rm \cQ}}
\def\rR {{\scriptscriptstyle \rm \cR}}

\def\cQb{{\cal \Qb}}
\def\cRb{{\cal \Rb}}
\def\cWb{{\cal \Wb}}

\def\fd {{\rm N}}
\def\afd {{\overline{\rm N}}}

\def \II {I\hspace{-.1em}I\hspace{.1em}}
\def \IIA {\mbox{\II A\hspace{.2em}}}
\def \IIB {\mbox{\II B\hspace{.2em}}}
\def \gs {g^s}
\def \ls {\lambda^s}

\def \I {{\cal I}}
\def \qs {q\hspace{-.53em}/\hspace{.15em}}
\def \ks {k\hspace{-.53em}/\hspace{.15em}}
\def \YM {{\mbox{\tiny YM}}}
\def \gym {g_{\YM}}

\def \Lc {\L_c}
\def\IR{\relax{\rm I\kern-.18em R}}
\def \id {{\bf 1}}

\def\cci{\ell}
\def\ccj{\ell'}

\author{Chong-Sun Chu  \\  
Centre for Particle Theory
and Department of Mathematics, 
Durham University, Durham, DH1 3LE, UK \\
E-mail:  
\email{chong-sun.chu@durham.ac.uk} }

\title {Cartan-Weyl 3-algebras and the BLG Theory I:\\
Classification of Cartan-Weyl 3-algebras }

\abstract{ 
As Lie algebras of compact connected Lie
groups, semisimple Lie algebras have wide applications in
the description of continuous symmetries of physical systems. Mathematically,
semisimple Lie algebra admits a Cartan-Weyl basis of generators which 
consists of a Cartan subalgebra of mutually commuting generators $H_I$ 
and a number of step generators $E^\a$ that are characterized by a root 
space of
non-degenerate one-forms  $\a$.
This simple decomposition in terms of the root space 
allows for a complete classification of semisimple
Lie algebras. In this paper, we introduce the analogous concept of  
a Cartan-Weyl Lie 3-algebra. We analyze their
structure and obtain  a complete classification of them. Many known 
examples of metric Lie 3-algebras (e.g. the Lorentzian 3-algebras) 
are special cases of the Cartan-Weyl 3-algebras. 
Due to their elegant and simple structure, 
we speculate  that Cartan-Weyl 3-algebras may be useful 
for describing some kinds of
generalized symmetries. As an application, 
we consider their use in the 
Bagger-Lambert-Gustavsson (BLG) theory.
}
 
\preprint{DCPT-10/05}
\keywords{D-Branes, M-Theory, Gauge symmetry, Lie $n$-algebra}

\begin{document}

\section{Introduction}

The analysis of the gravitational thermodynamics  has suggested
that the entropy of a large number of $N$ coincident branes should obey
a power law scaling $N^2, N^{3/2}, N^3$ for D-branes, M2-branes and
M5-branes respectively \cite{KT}. For coincident D-branes, the $U(1)$ gauge
symmetry for individual D-brane is enhanced to a $U(N)$ gauge symmetry,
and the $N^2$ dependence of the entropy is nicely accounted for by the
$N^2$ degrees of freedom present in the supersymmetric $U(N)$ Yang-Mills
description of the coincident D-branes. The situation is much less clear
for multiple M2 or M5-branes. Some of the outstanding questions are, 
for example: How does the gauge symmetry get enhanced for coincident M-branes? 
What is the appropriate mathematical description?

Recently a new class of (2+1)-dimensional superconformal field theories
with maximal $\cN=8$ supersymmetry has been constructed 
by Bagger and Lambert \cite{BL1,BL2,BL3}, 
and separately by Gustavsson \cite{G1}. These field theories have 
been proposed as the low energy effective field theories for
coincident M2-branes and thus  provide a nonperturbative description of
M-theory on $AdS_4 \times S^7$ according to the AdS/CFT correspondence 
\cite{adscft}.
Another proposal is due to Aharony, Bergman, Jafferis and Maldacena 
\cite{ABJM} which proposed a certain $\cN=6$ superconformal 
Chern-Simons-matter theory as  the theory describing the
worldvolume of multiple M2-branes at low energies \cite{adscft-ABJM}. 
In this description,
a total of $\cN=6$ supersymmetry is explicitly realized \cite{susy}.

One of the most exciting features of the BLG theory is 
the use of a new mathematical
structure, a metric Lie 3-algebra, in its description of the gauge symmetries
of the multiple M2-branes. It has also been suggested that
Lie 3-algebras also play a crucial role in the description of multiple M5-branes
 \cite{CS1,CS2,LP}.

In the case of D-branes, the worldvolume theory 
carries a gauge symmetry that
is described by a semisimple Lie algebra. Indeed
the class of semisimple Lie algebras is 
distinguished in the theory of Lie algebras.
Physically,
semisimple Lie algebras are the Lie algebras of compact connected Lie
groups, one that are universally used to describe the continuous
symmetries of physical systems. 
Mathematically, semisimple Lie
algebras are completely classified and are fully understood. 
In particular,
semisimple Lie algebras admit an elegant and very simple
basis of generators called the {\it Cartan-Weyl basis}. See
equations \eq{metric-cA} and \eq{cw2-1} below. 
Motivated by the simplicity of the structure of a
Cartan-Weyl basis of generators for a semisimple Lie algebra, 
one can introduce the notion
of a Cartan-Weyl basis for a metric Lie 3-algebra. A Cartan-Weyl basis 
consists of the generators 
${H_I, E^\a}$ where $H_I$ ($I =1, \cdots, N$ for some
$N$) form a Cartan subalgebra $\cH$ of
the Lie 3-algebra and $E^\a$ are labelled by {\it roots} $\a$, which are
linear functions on $\cH^{\wedge 2}$. 
The generators  satisfy the Lie 3-brackets
\be \label{ccw1}
[H_I,H_J,H_K]= 0
\ee
and 
\bea \label{ccw2}
\; [H_I, H_J, E^\a]& =& \a_{IJ} E^\a, \nn\\
\; [H_I,E^\a, E^\b] &=&  \begin{cases}
\a_{IK} g^{KL} H_L, & \mbox{if $\a+ \b =0$}, \\
g_I(\a,\b) E^{\a+\b}, & \mbox{if $\a+ \b \neq 0$ is a root},\\
0,  & \mbox{if $\a+ \b$ is not a root}, 
\end{cases}
\\
\; [E^\a,E^\b, E^\g] &=&  \begin{cases}
-g_K(\a,\b) g^{KL} H_L,  & \mbox{if $\a+ \b + \g =0$}, \\
c(\a,\b,\g) E^{\a+\b+\g},  & \mbox{if $\a+ \b + \g \neq 0$ a root},\\
0,  & \mbox{if $\a+ \b + \g$ is not a root};
\end{cases}
\nn
\eea
and they are normalized such that 
\be
\la E^\a, E^\b \ra = \d^{\a+\b}, \quad \la H_I, E^\a \ra =0, \quad 
\la H_I, H_J \ra  = g_{IJ}.
\ee
Here $\la \cdot , \cdot\ra$ is the metric and the part 
$g_{IJ}$ is assumed to be invertible  with the inverse $g^{IJ}$.
We will call a metric
Lie 3-algebra a {\it Cartan-Weyl 3-algebra}
if it admits such a {\it Cartan-Weyl basis}. 
It is natural to speculate that a 
classification theorem may be obtained for the 
Cartan-Weyl 3-algebras.

Due to their similarity in structure to semisimple Lie algebras,
it is  natural to speculate  that Cartan-Weyl 3-algebras
may also play a role in the description of certain yet to be discovered
generalized symmetries of Nature.  
The understanding of the structure and the classification of Cartan-Weyl
3-algebras are therefore potentially important.
To achieve these goals is the main motivation of this paper.
Once this is achieved, it is natural to try to see whether and how 
this kind of Lie 3-algebras is useful for the BLG theory. This is another 
motivation of this work.

The organisation of the paper is as follows.
In section 2.1, we review some basic facts about Lie algebras. In
particular we recall the definition of a Cartan-Weyl basis.
In section 2.2, we  motivate and introduce the definition of 
a Cartan-Weyl basis for a
metric Lie 3-algebra. The consistency conditions arising from the fundamental
identity are analysed in details in the appendix. 
The resulting conditions that the two-form roots $\a_{IJ}$ and 
the coefficients $g_I(\a,\b)$ 
and $c(\a,\b,\g)$  have to satisfy are
summarized in section 2.3. In section 3, these conditions are  solved
fully and a complete classification of Cartan-Weyl 3-algebras is obtained. 
In section 4, we 
consider the embedding of $\cA_4$ 
\footnote{The Lie $n$-algebra 
$\cA_{p,q}$ is a metric Lie $n$-algebra with signature $(p,q)$, $p+q
=n+1$. 
It has $n+1$ generators $e_i$, $i =1, \cdots, n+1$ and is defined
by  the metric
\be
\langle e_i, e_j \rangle = \ve_i \d_{ij}
\ee
and the $n$-bracket relations 
\be
[e_1, \cdots, \hat{e_i}, \cdots, e_{n+1}] = (-1)^i \ve_i e_i.
\ee 
The signs $\ve_i$ are
given by $(+ \cdots +)$ for $\cA_{0,n+1} := \cA_{n+1}$, 
$(-+ \cdots ++)$ for $\cA_{1,n}$,   $(--+\cdots +)$ for $\cA_{2,n-1}$ etc.
}
and
show that  Cartan-Weyl
3-algebras do not contain $\cA_4$ as a subalgebra in general. 
This implies that a BLG
theory that is based on a Cartan-Weyl 3-algebra cannot contain fuzzy
$S^3$ in its description, at least not semiclassically.  Section 5 
contains some further discussions. The detailed analysis of the consistency
conditions is performed in the appendix A. 
 
\section{Cartan-Weyl 3-algebras and Consistency Conditions}
\label{ccw}

\subsection{Cartan-Weyl basis for a semisimple Lie algebra}
 
We start with some basic definitions about Lie algebras. 
Let $\mg$ be a Lie algebra over the field $\CC$. 
A subspace $\ma$ is a {\it subalgebra} if
$[\ma,\ma] \subset \ma$.  A subspace $I$ is an {\it ideal} if $[I,\mg]
\subset I$. 
Suppose that $I_1, I_2$ are ideals of a Lie algebra $\mg$, 
then $[I_1,I_2]$ is also an ideal of $\mg$.
In particular, $\mg$ has the following two series of ideals: 
\be
\mg \supset \mg^1 \supset \mg^2 \cdots,
\ee
which is called the {\it descending central series}of $\mg$; and
\be
\mg \supset \mg^{(1)} \supset \mg^{(2)} \cdots,
\ee
which is called the  {\it derived series} of $\mg$. 
Here 
\be
\mg^{0} := \mg, \quad \mg^{n} : =[\mg^{n-1},\mg]
\ee
and
\be
\mg^{(0)} := \mg, \quad \mg^{(n)} : =[\mg^{(n-1)},\mg^{(n-1)}]
\ee
for $n \in \NN$.
A Lie algebra
$\mg$ is {\it nilpotent} if $\mg^{n} =0$ for some $n$. It 
is {\it solvable} if $\mg^{(n)} =0$ for some $n$.
Obviously every nilpotent Lie algebra is also solvable.

A Lie algebra is {\it simple} if it has no proper ideals other than
itself or 0.
The union of all solvable ideal of $\mg$ is also a solvable ideal of
$\mg$, called the radical $\mR(\mg)$  of $\mg$.  A Lie algebra is {\it
semisimple} if its radical is 0. 
\if
The significance of semisimplicity is due to the {\it Levi
decomposition} which states that 
\begin{thm}{(Levi decomposition)} Every finite dimensional Lie algebra $\mg$
admits a Levi subalgebra $\ml$, i.e. there exists 
a semisimple subalgebra $\ml$ of $\mg$ such
that $\mg$ is a semidirect sum of its radical $\mR(\mg)$ and $\ml$
\be
\mg = \mR(\mg) \oplus \ml
\ee
\end{thm}
\fi
As a result, a semisimple Lie algebra is given by a direct sum of
simple Lie algebras and Abelian ones.

In the theory of Lie algebra, a very useful device  is the {\it Killing metric}
which can be defined using the
adjoint representation ${\rm ad} : \mg \to \mg$ by
\be
\k (x,y) = \tr \big( ({\rm ad}\, x)({\rm ad}\, y) \big), \quad \mbox{for
  $x,y \in \mg$ }.
\ee
The Killing metric is invariant:
\be
\k ([x,z],y) +\k(x,[y,z]) =0, \quad \mbox{for all $x,y,z \in \mg$}.
\ee
Killing metric plays an important role in characterizing Lie algebra. In
fact,
\begin{thm}{(Cartan criteria of semisimplicity)} \label{C1}
A Lie algebra $\mg$ is semisimple iff the Killing metric $\k$ is non-degenerate.
\end{thm}

\begin{thm}{(Cartan criteria of solvability)} \label{C2}
A Lie algebra $\mg$ is solvable iff the Killing metric  $\k([x,y],z)
=0$ for all $x,y,z \in \mg$.
\end{thm}

It is instructive  to recall
the reason why semisimple Lie algebras play a fundamental role  in
physics.  In
fact the classification theorem of Lie groups
states that any compact, connected
Lie group is a product of a finite abelian group and
simple Lie groups \footnote{We also recall that a simple Lie group
is isomorphic to exactly one of the $SU(n), n\geq 3;  Sp(n) ,n\geq
1;  Spin(n), n\geq 7;  G_2, F_4, E_6, E_7, E_8$.}.
By modelling the continuous symmetries that appears in nature
by compact connected Lie groups, one thus arrive at the semisimple Lie
algebras naturally.
 
An important result in the theory of Lie algebra  
is that a semisimple Lie algebra admits a
very nice basis of generators called the Cartan-Weyl basis.
A Cartan-Weyl basis consists of a set of generators $H_I$ from the 
Cartan subalgebra and
a set of step generators $E^\a$  that are labelled by a vector
$\a=(\a_I)$ called the root. 
In general any two Cartan subalgebras are conjugate  relative to the group
of special automorphisms generated by the exponents of nilpotent inner 
derivations \cite{jacobson,OV}.
A special feature of the Cartan-Weyl
basis is that the Cartan subalgebra is Abelian and 
the roots are  non-degenerate.  
In this basis, the Killing metric reads
\be \label{metric-cA}
\la E^\a, E^\b \ra = \d_{\a+\b}, \quad \la E^\a, H^I \ra =0, 
\quad g_{IJ} := \langle H_I,H_J \rangle,
\ee
where  $g_{IJ}$, the restriction of the Killing metric on the Cartan 
subalgebra, is non-degenerate for a semisimple Lie algebra.
The Lie brackets take the form
\bea \label{cw2-1}
[H_I,H_J] &=& 0, \nn\\
\;  [ H_I, E^\a ] &=& \a_I E^\a, \nn\\
\; [E^\a, E^\b] &=& \begin{cases}  
0, & \mbox{if $\a+\b \neq 0$ not  a root},\\
c(\a,\b) E^{\a+ \b}, & \mbox{if $\a+\b \neq 0$ is a root},\\
- \a \cdot H & \mbox{if $\a+\b =0$},
\end{cases} 
\eea
where $\a \cdot H = \a_I g^{IJ} H_J$ and  $g^{IJ}$ is the inverse of
$g_{IJ}$.
Here we have used the invariance of the metric in 
deriving the relation for $[E^\a,E^{-\a}]$. The Cartan-Weyl basis is
specified by the system of roots and the coefficient
$c(\a,\b)$. They can be  solved and gives a complete classification of
semisimple Lie algebras. 

The proof of the existence of a Cartan-Weyl basis for a semisimple Lie
algebra rests on the theory of root space decomposition  for Lie
algebras, see for example, \cite{jacobson,OV}. We have also included a
proof in the companion paper \cite{cat2}, highlighting the most important ideas
involved so as to explain the conditions that are needed to establish the
existence of a {\it generalized Cartan-Weyl basis}
for a Lie 3-algebra. 
Here let us take an elementary and more direct approach 
to explain  
the assumptions involved that lead to 
the existence of a generalized Cartan-Weyl basis. This 
exercise will also be useful for 
a heuristic understanding the conditions that lead to 
the existence of a similar Cartan-Weyl basis for a Lie 3-algebra.

Given a Lie algebra, let us start by looking
for a maximal set of commuting generators $H_I$, $I =1, \cdots,
N$. Denote the rest of the generators by $E^\a$ where $\a$ is just a
label at this point. 
Denote also the set of all generators by $Y^a = \{ H_I,
E^\a\}$. Most generally we have
\be \label{eigen}
[H_I, Y^a] = \psi_I{}^a{}_b Y^b,
\ee
where $\psi_I= (\psi_I){}^a{}_b$ is a constant matrix. Jacobi identity
$[H_I,[H_J, Y^a]] + \cdots =0 $ implies that
\be
[\psi_I, \psi_J] = 0
\ee
and so we can diagonalize $\psi_I$'s simultaneously. If one assumes that 
all the
eigenvalues are non-degenerate, one obtains 
$\psi_I{}^a{}_b = \d^a{}_b \psi_{Ia}$. Obviously $\psi_{Ia} =0$ for
$a=K$. Therefore
\be 
[H_I, E^\a] = \a_I E^\a,
\ee
where we have denoted $\psi_{I \a} = \a_I$.
Next using the invariance of the metric $\d_{H_I}\la E^\a, E^\b \ra =0$
and $\d_{H_I}\la H_I, E^\a \ra=0$, one obtains immediately 
\be
\la E^\a, E^\b \ra \propto \d^{\a+\b}, \quad \la H_I, E^\a \ra =0.
\ee
Normalizing the generators suitably and assuming that the metric is 
non-degenerate when restricted to the $H$'s, 
we obtain the metric \eq{metric-cA}. Next we note
that the Jacobi identity gives $ [H_I,[E^\a,E^\b]] = (\a+\b)_I
[E^\a,E^\b]$. Therefore, using also the invariance of the metric, one sees 
that
$[E^\a, E^\b] $ takes the form as in \eq{cw2-1}. This is precisely 
the Cartan-Weyl basis. 

What we
see from this analysis is that the existence of a Cartan-Weyl basis is
equivalent to the assumption that the solutions of the 
``eigenvalue equation'' \eq{eigen}  are non-degenerate; and that the
restriction of the metric to the $H$'s
is non-degenerate. This is also 
equivalent to the assumption of semisimplicity.
For Lie 3-algebras, we will now show that a Cartan-Weyl basis exists with
similar assumptions of non-degeneracy.

\subsection{Cartan-Weyl 3-algebras and the Cartan-Weyl basis}

A Lie $n$-algebra $\cA$ \cite{filippov,K1} is a linear space over a 
field $\FF$ on which defined is
a multilinear $n$-bracket operation which is skew-symmetric and
satisfies the fundamental identity:
\be
[ [b_1, \cdots, b_n], a_1, \cdots, a_{n-1}] = \sum_{i=1}^n 
[ b_1, \cdots, [b_i, a_1, \cdots, a_{n-1}], \cdots, b_n]
\ee
for all $a_i, b_j \in \cA$. 
For $n=2$, we get back to the usual 
Lie algebra and the fundamental identity coincides
with the   Jacobi identity. We will be concerned with 
real Lie $n$-algebras in this paper.

The fundamental identity guarantees that the transformation
$\d: \cA \rightarrow \cA$ defined by
\be
\d_{(a_1,\cdots,a_{n-1})} f := [f, a_1,\cdots, a_{n-1}]
\ee
is a derivation of the Lie $n$-algebra:
\be
\d [b_1, \cdots, b_n] = [\d b_1, \cdots,b_n]+ \cdots + [b_1,\cdots,\d b_n].
\ee 
The map $\d$ is parametrized by a skew-symmetric  
collection of $n-1$ elements  $(a)=(a_1, \cdots, a_{n-1})$ and is a 
natrual generalization of
the usual gauge transformation $\d_a f:= [f,a]$ defined for Lie algebra. 

A {\it metric} $\la\cdot , \cdot \ra : \cA \otimes \cA \rightarrow \RR$ 
is a symmetric bilinear form on $\cA$ which is also invariant in the
sense that
\be
\d_{(a)}\la f,g \ra = 0,
\ee
i.e.
\be
\la [f,(a)], g\ra + \la f, [g,(a)]\ra =0,
\ee
for any $f,g \in \cA$ and $(a) \in \cA^{\wedge (n-1)}$.
In addition to a metric, it is also natural to introduce the notion of
an
{\it invariant form}. An invariant form on  a Lie $n$-algebra $\cA$ is a
multilinear function $F$ on $\cA^{\wedge (n-1)} \times \cA^{\wedge (n-1)}$
which satisfies the invariance condition:
\bea \label{inv-form}
\sum_{i=1}^{n-1}
&& F(a_1, \cdots , a_i R_{(c)}, \cdots, a_{n-1}; b_1, \cdots, b_{n-1}) \nn\\
&& \qquad 
+ F( a_1, \cdots, a_{n-1};b_1, \cdots, b_i R_{(c)}, \cdots, b_{n-1}) =0,
\eea
where $R_{(c)}$ is the right multiplication 
$a R_{(c)} :=[a,c_1, \cdots, c_{n-1}]$. 
We say an invariant form is non-degenerate if $F(a_1, \cdots, a_n ;b_1,
\cdots, b_{n-1}, x ) =0$ for all $a_1, \cdots, a_n, b_1, \cdots, b_{n-1}
\in \cA$ implies that $x=0$. 
In this paper we will be mainly interested in Lie 3-algebra with an invariant
non-degenerate metric.  We will refer to this as a metric Lie 3-algebra.

In analogy to the Lie algebra case, 
one may introduce the notion of a
Cartan-Weyl basis for a Lie 3-algebra as follows. Let $\cA$ be a metric Lie
3-algebra, we call $\cA$ a {\it Cartan-Weyl 3-algebra} if the algebra admits
a basis of generators $H_I$, $I =1, \cdots, N \geq 2$ and $E^\a$, $\a
= (\a_{IJ})$, with the metric 
\be  \label{metric3-eeh}
\la E^\a, E^\b \ra = \d_{\a+\b},\quad \la E^\a, H_I \ra =0,
 \quad g_{IJ}:= \langle  H_I, H_J \rangle \;\;\mbox{non-degenerate}, 
\ee 
and the 3-brackets: 
\bea 
\; [H_I, H_J, H_K] &=& 0, \label{cw31} \\
\; [H_I, H_J, E^\a]& =& \a_{IJ} E^\a, \label{cw32}
\eea
\begin{subnumcases}{[H_I,E^\a, E^\b] =}
\a_{IK} g^{KL} H_L, & \mbox{if $\a+ \b =0$,} \label{cw33a}\\
g_I(\a,\b) E^{\a+\b}, & \mbox{if $\a+ \b \neq 0$ is a root,} \label{cw33b}\\
0,  & \mbox{if $\a+ \b$ is not a root,} \label{cw33c}
\end{subnumcases}
\begin{subnumcases}{[E^\a,E^\b, E^\g] =}
- g_K(\a,\b) g^{KL} H_L,  & \mbox{if $\a+ \b + \g =0$,} \label{cw34a} \\
c(\a,\b,\g) E^{\a+\b+\g},  & \mbox{if $\a+ \b + \g \neq 0$ a root,} 
 \label{cw34b}\\
0,  & \mbox{if $\a+ \b + \g$ is not a root.}   \label{cw34c}
\end{subnumcases}
In writing these 3-bracket relations, we have used the invariance of the 
metric to relate \eq{cw33a} with \eq{cw32} and \eq{cw34a} with \eq{cw33b}.
The number of independent $H$'s will be called the {\it rank} of the
Cartan-Weyl 3-algebra and the set of generators $\{H_I, E^\a\}$
will be called a {\it Cartan-Weyl basis}. Similarly one can define a 
Cartan-Weyl Lie $n$-algebra.

It is natural to ask if semisimplicity is again the right condition to
guarantee the existence of a Cartan-Weyl basis for a Lie 3-algebra
(or Lie $n$-algebra in general). It turns out that 
there is a number of ways one can generalize the concept of
semisimplicity of Lie algebra to the higher
Lie $n$-algebra, and so the answer is not obvious.
In the paper \cite{cat2}, we will investigate this question and 
find out what are the precise mathematical
conditions for a Lie 3-algebra to be a (generalized) Cartan-Weyl
3-algebra.
 
Without entering into these more abstract discussions, it is possible
to state and explain quite explicitly what are the conditions 
that lead to the existence of a  Cartan-Weyl basis. In analogy to
the discussion for  Lie algebras,  let us start by saying we have a
maximal set of ``commuting'' generators 
in the sense that 
\be
[H_I,H_J,H_K] =0, \quad I,J,K = 1,\cdots, N. 
\ee
Denote the rest of the generators by $E^\a$ and the set of all
generators by $Y^a = \{ H_I, E^\a\}$. Most generally we have
\be \label{eigen2}
[H_I, H_J, Y^a] = \psi_{IJ}{}^a{}_b Y^b,
\ee
where $\psi_{IJ}= (\psi_{IJ}){}^a{}_b$ is a constant
matrix. Fundamental identity $[H_K,H_L,[H_I,H_J,Y^a]] = \cdots $
implies that 
\be
[\psi_{IJ}, \psi_{KL}] = 0
\ee
and so we can diagonalize $\psi_{IJ}$'s simultaneously.
If one assume that all the eigenvalues are non-degenerate, then we have 
$\psi_{IJ}{}^a{}_b = \d^a{}_b \psi_{IJ\,a}$. Obviously $\psi_{IJ\,a} =0$ for
$a=K$. Denoting $\psi_{IJ\, \a} = \a_{IJ}$, we have
\be \label{eigen1}
[H_I, H_J, E^\a] = \a_{IJ} E^\a
\ee
and so the $E$-type generators are parametrized by a two-form
$\a_{IJ}$. 
Next using the invariance of the metric $\d_{(H_I,E^\a)}\la E^\b, H_J \ra =0$
and $\d_{(H_I,H_J)}\la E^\a, H_K \ra=0$, we obtain immediately 
\be
\la E^\a, E^\b \ra \propto \d^{\a+\b}, \quad \la H_I, E^\a \ra =0.
\ee
Normalizing suitably the generators 
and assuming that the metric is non-degenerate
when restricted to the $H_I$'s, we get precisely the metric 
\eq{metric3-eeh}. 
Next use the fundamental identity $[H_J,H_K,[H_I, E^\a,E^\b]] = \cdots$, 
we obtain 
$[H_J,H_K,[H_I, E^\a,E^\b]] = (\a+\b)_{JK} [H_I, E^\a,E^\b]
$. Together with the assumption that the $H_I$'s  forms a maximal
set of commuting generators, we obtain 
$[H_I, E^\a,E^\b] =c_I(\a){}^J H_J$  for $\a+ \b=0$. Invariance of the
metric then implies that it is precisely of the form \eq{cw33a}.
For $\a+\b \neq 0$, if we denote
\be
[H_I, E^\a,E^\b] = \l_I{}^{\a\b}{}_a Y^a,
\ee
then the fundamental identity $[[H_I,H_J,H_K],E^\a,E^\b] = \cdots$ gives
\be \label{lc1}
\sum_{ \mbox{$(I,J,K)$ cyclic}}
 \l_I{}^{\a\b}{}_\g \g_{JK} =0.  
\ee
On the other hand, the invariance of the metric 
$\d_{(H_I,E^\a)} \la E^\b,E^\g \ra =0$ implies
that
\be \label{lc2}
 \l_I{}^{\a\b\g} = \l_I{}^{[\a\b\g]} 
\ee
is completely antisymmetric in $\a,\b, \g$. Here we have used the metric
$\la E^\a,E^\b \ra = \d^{\a+\b} $ to raise an indices of $\l$. 
Using \eq{lc1} and \eq{lc2} and the fundamental identity
$[[H_i,H_j,E^\a],H_K,E^\b] = \cdots$, one can easily deduce that 
$\l_I{}^{\a\b\g} = 0$ unless $\a+\b+\g =0$. Therefore 
$\l_I{}^{\a\b\g}$ takes the form
\be
\l_I{}^{\a\b\g} = g_I(\a,\b), \quad \mbox{for $\a+\b+\g =0$}
\ee 
and so we obtain the 3-bracket \eq{cw33b}.
Finally, assume that
\be
[E^\a,E^\b, E^\g] = \o^{\a\b\g}{}_a Y^a.
\ee 
By considering the
fundamental identity $[[E^\a,E^\b,E^\g],H_J,H_K] = \cdots$, one can show that
\be \label{omega}
(\a+\b+\g+\d) \o^{\a\b\g\d} =0, \quad 
(\a+\b+\g)  \o^{\a\b\g L} =0.
\ee
The first condition implies that $\o^{\a\b\g\d} =0$ unless $\a+\b+\g+\d =0$. 
In this case, by denoting $\o^{\a\b\g\d} = c(\a,\b,\g)$, we arrive at 
the 3-bracket \eq{cw34b} after using the second condition of
\eq{omega}. The second condition of \eq{omega}
gives precisely the 3-brackets \eq{cw34a} after using 
the invariance of the metric. 

All in all, this analysis demonstrates that 
the existence of a Cartan-Weyl basis for a Lie 3-algebra is
equivalent to the requirement that 
``eigenvalue equation'' \eq{eigen2}  has non-degenerate solutions; 
and that the
restriction of the metric to the $H$'s
is non-degenerate.

\subsection{Consequences of the consistency conditions}

A Cartan-Weyl 3-algebra is specified by the data:
the roots $\a_{IJ}$ and the structural constants 
$g_I(\a,\b)$ and $c(\a,\b,\g)$. 
Remarkably, the consistency conditions arising from the fundamental 
identities are so strong that one can solve them exactly, giving a  
complete classification of 
Cartan-Weyl 3-algebras.

The consistency conditions that follow from the fundamental
identities
are listed and analyzed in details in the
appendix. It turns out  that we have generally  
\be
\boxed{
c(\a,\b,\g) =0.
}
\ee
As for the roots, they generally have to satisfy the condition \eq{cond1}.
For a Cartan-Weyl
3-algebra with a
single pair of roots $\pm \a$,  \eq{cond1} is the only condition to be
satisfied. 
For Cartan-Weyl 3-algebra with more than one pair of roots,  
the root space can generally be decomposed into a number
of (say $M$) components: $\D_\cA(\cH) = \oplus_{\cci=1}^M \Omega_\cci$, 
where there is a null 
vector $\ph^{(\cci)}$ associated with each $\Omega_\cci$. The roots in each
$\Omega_\cci$ can be decomposed in the form
\be \label{cond-ap}
\boxed{
\a = \ph^{(\cci)} \wedge \ah^{(\cci)}, \quad \a \in \Omega_\cci.
}
\ee
The corresponding
one-form parts $\ah^{(\cci)}$ form the root system of a semisimple Lie
algebra $\mg^{(\cci)}$.
Moreover $\ph^{(\cci)}$ and $\ah^{(\ccj)}$ satisfy the conditions
\be \label{cond6ap}
\boxed{
\ph^{(\cci)} \cdot \ph^{(\ccj)} =0, 
\quad \mbox{for all $\cci,\ccj$},
}
\ee
\be \label{cond6p}
\boxed{\ph^{(\cci)} \cdot \ah^{(\ccj)} =0,
\quad \mbox{for all $\cci,\ccj$},
} 
\ee
\be\label{cond7p} \boxed{
\ah^{(\cci)} \cdot \bh^{(\ccj)} =0, \quad \mbox{for $\cci \neq \ccj$},
}
\ee
where the dot product is taken with respect to the metric $g_{IJ}$ of the 
Lie 3-algebra.
As for $g_I(\a,\b)$, we find
\be \label{cond-gI}
\boxed{
g_I (\a,\b)  = \begin{cases}
\ph_I^{(\cci)} c^{(\cci)}(\ah^{(\cci)},\bh^{(\cci)}), & \quad \mbox{for $\a,\b \in
\Omega_\cci$}, \\
0, & \quad \mbox{otherwise}.
\end{cases}
}
\ee
where the coefficients $c^{(\cci)}$ 
specifies the ``$[E,E]=E$''  type brackets of 
the semisimple Lie algebra $\mg^{(\cci)}$
as in \eq{lie-cw}. 
\if
Finally the coefficient $c(\a,\b)$ is given by 
\be \label{cond-cp}
\boxed{
c(\a,\b) = \begin{cases}
c^{(\cci)}(\ah^{(\cci)},\bh^{(\cci)}),
& \mbox{$\a,\b \in \Omega_\cci$, }
 \\
0, & \mbox{otherwise}.
\end{cases}
}
\ee
\fi

The construction and classification of Cartan-Weyl
3-algebras has thus been reduced  to the problem of constructing 
the null vectors $\ph^{(\cci)}$ and roots $\ah^{(\cci)}$ that satisfy the
conditions  \eq{cond6ap}-\eq{cond7p}. This can be done fully 
and explicitly. This will be our task in the next section. 

\section{Explicit Classification of Cartan-Weyl 3-algebras}\label{explicit}

Let us first consider the case of a 
rank $N$ Cartan-Weyl 3-algebra with only a
single pair of roots $\pm \a$.
In this case, the condition \eq{cond1} is the only solution to be
satisfied. The general solution of it is
\be
\a = \a_1 \wedge \a_2,
\ee 
where $\a_1, \a_2$ are linearly independent. Given $\a_1, \a_2$, one can
find a basis of $N$ vectors $p_1, p_2, q_3, \cdots q_N$ 
such that  $q_l \cdot \a_1 = q_l \cdot \a_2 =0$ and $p_1$, $p_2$ lies
in the plane spanned by $\a_1$ and $\a_2$. It follows immediately
that
\bea
\; && [ q_l\cdot H, H_I, E^{\pm\a}] = 0, \nn\\
\; && [ q_l\cdot H, E^\a, E^{-\a}] =0, \quad l =3, \cdots, N. 
\eea 
Thus the $N-2$ generators $q_l\cdot H$ are central in the algebra and
the nontrivial part of the Lie 3-algebra 
has  only 4 generators $\{ p_1 \cdot H, p_2 \cdot
H, E^\a, E^{-\a} \}$. Depending on the signature of the vector space
spanned by the vectors $p_1, p_2$, this 3-algebra is isomorphic to
$\cA_{0,4}, \cA_{1,3}$ or $\cA_{2,2}$.

Therefore  let us consider the general case where the  
Cartan-Weyl 3-algebra
has more than a single pair of root. 
Since our results given in the section 2.2 relies on the
existence of the null vectors $\ph^{(\cci)}$, therefore it is natural to
classify the  
Cartan-Weyl 3-algebra according to the number of negative 
eigenvalues of the metric $g_{IJ} := \la H_I, H_J \ra$. With a Hermitian
structure 
\be 
(H_I)^\dag =H_I, \quad (E^\a)^\dag = E^{-\a},
\ee
this is also the same as the index of
the metric over the whole Lie 3-algebra $\cA$.

\subsection{Index  0 and 1}

For index 0,   Cartan-Weyl 3-algebra 
is possible only if the 3-algebra has
a single pair of root. In this case the 3-algebra is isomorphic to
$\cA_{0,4}$ plus a number of central elements. This is decomposable.
Obviously it is sufficient to classify indecomposable   Cartan-Weyl
3-algebra. In the following we will
consider only indecomposable   Cartan-Weyl 3-algebra.

Next consider the case of index 1. Let us choose a basis of the Cartan
subalgebra as $\{H_I\} = \{ H_\Ih,H_a\}$, $\Ih=1, \cdots, \frN$, $ a=1,2$, 
such that   the metric 
takes the form 
\be 
g_{I J} = 
\begin{pmat} ( {|.} )
g_{\Ih \Jh} &   &  \cr \- 
 & 1 & \cr
 & & -1 \cr
\end{pmat},
\ee
where  $g_{\Ih \Jh}$ is Euclidean. Here we have the rank  $N= \frN+2$.
 
\noindent \underline{root system}

To construct a    Cartan-Weyl 3-algebra, we
need to pick $\ah^{(\cci)}$ and null vectors $\ph^{(\cci)}$ 
such that \eq{cond6ap}-\eq{cond7p} are satisfied. Knowing already that
$\ah^{(\cci)}$ are roots of a Lie algebra, the most general
solution is to take $(\ah_I^{(\cci)}) = (\ah_\Ih^{(\cci)}, 0, 0)$ and to take
the null vectors in the orthogonal complements $\ph^{(\cci)} = ({\bf 0},
\cdot, \cdot)$. There can be (up to overall normalization) 
two choices of such null vectors. But since they cannot be orthogonal
to each other, it means we can only use one  null vector in the index
1 case. Without loss of generality, let us take
\be
\ph= ({\bf 0}, 1, 1)/\sqrt{2},
\ee
and the roots of the   Cartan-Weyl 3-algebra are of the form
\be
\a = \ph \wedge \ah, \quad \mbox{where}\quad (\ah_I)= (\ah_\Ih,0,0) .
\ee
Explicitly, it has the nonvanishing components $\a_{a \Ih}= \ph_a \ah_\Ih$. 

\noindent\underline{3-brackets}

It is convenient to reorganize the basis of Cartan generators as
\be
H_I = \{ H_\Ih, v, u \},
\ee
where
\be
v:= \ph \cdot H, \quad u:= \pt \cdot H
\ee
with 
\be
\pt := ({\bf 0}, 1, -1)/\sqrt{2}. 
\ee
It is 
\be \label{i1-0}
\la v,v\ra = \la u, u \ra =0 \quad \mbox{and $\la v,u\ra =\ph \cdot \pt
=1$}.
\ee  
Since the ``step''
generators $E^\a$ are characterized by the one-form part $\ah$, 
it can be written as $E^\a = E^\ah$.  
As a vector space,
the Cartan-Weyl 3-algebra of index 1 is given by
\be
\cA = \mg \oplus \CC(u,v).
\ee
The   Cartan-Weyl 3-algebra relations read
\bea
\; &&[v, \cdot, \cdot]= 0, \label{i1-1}\\
\;&& [u,H_\Ih, E^\ah] = \ah_\Ih E^\ah, \label{i1-2}\\
\; && [u, E^\ah, E^\bh] = \begin{cases}
\ah_\Ih g^{\Ih\Jh} H_\Jh, & \ah+\bh=0, \\
c(\ah,\bh)E^{\ah+\bh}, &  \mbox{$\ah+\bh$ nonzero roots}, 
\end{cases} \label{i1-3}\\
\; && [E^\ah, E^\bh, E^\gh] = \begin{cases}
-c(\ah,\bh) v, & \ah+\bh+\gh =0, \\
0, & \mbox{otherwise}.
\end{cases} \label{i1-4}
\eea
The relations \eq{i1-1} - \eq{i1-4} can be summarized as
\bea \label{i1-5}
\; && [v, \cdot, \cdot]= 0, \nn \\
\; && [u, g_1,g_2]  = [g_1,g_2]_\mg, \\
\; && [g_1,g_2,g_3] = -\la [g_1,g_2], g_3 \ra_\mg v , \nn 
\eea
where $g_i \in \CC(H_I, E^\ah):=\mg$ is a semisimple Lie algebra with
the  brackets $[\cdot,\cdot]_\mg$ 
and the metric   $\la \cdot , \cdot \ra_\mg$
as defined by \eq{metric-cA}.

We note that the Cartan-Weyl 3-algebra of index 1 
is precisely the same as  the
Lorentzian 3-algebra \cite{LOR}.

\subsection{Index 2}

For index 2, 
let us choose a basis of the Cartan generators as $\{H_I\}= \{H_\Ih, H_a
\}$, $\Ih = 1, \cdots, \frN$, $a=1, \cdots, 4$, 
such that the metric takes the form with nonzero entries
\be \label{m2}
\begin{pmat} ( {|.} )
g_{\Ih \Jh} &   &  & &	\cr\- 
 & 1 &   &  &         	\cr
 &   & -1&  & 		\cr  
 &   &   & 1& 		\cr
 &   &   &  & -1        \cr
\end{pmat},
\ee
where  $g_{\Ih \Jh}$ is Euclidean. In this case, the rank  $N= \frN+4$.

\noindent\underline{root system}

We can solve the conditions \eq{cond6p} by
taking $\a_I =(\a_\Ih; 0,0,0,0)$ and the null vectors in the
complementary part of the vector
space.   There are four such linearly independent  null vectors. 
One can choose the basis of null
vectors $\{ \ph^{(1)}, \ph^{(2)}, \pt^{(1)},\pt^{(2)} \}$ such that 
$\ph^{(\ri)} \cdot \ph^{(\rj)} = 0$,
$\pt^{(\ri)} \cdot \pt^{(\rj)} = 0$,
$\ph^{(\ri)} \cdot \pt^{(\rj)} = \d_{\ri\rj}$
\footnote{
For example, with respect to the
metric \eq{m2}, one can take 
$\ph^{(1)}= ({\bf 0}, 1,1,0,0)/\sqrt{2}$, 
$\ph^{(2)}= ({\bf 0}, 0,0, 1,1)/\sqrt{2}$,
$\pt^{(1)}= ({\bf 0}, 1,-1,0,0)/\sqrt{2}$, 
$\pt^{(2)}= ({\bf 0}, 0,0, 1,-1)/\sqrt{2}$. 
However we will not need
these explicit expressions.
}. 
Moreover since any linear combinations of the
$\ph$-vectors (or of the $\pt$-vectors) are still null, 
the most general set (up to trivial overall normalization factor) 
of null vectors which satisfy \eq{cond6ap} is thus given by
\be \label{p1s}
\ph^{(\l)} := \ph^{(1)} + \l \, \ph^{(2)},  
\ee 
where $\l$ is an arbitrary nonvanishing finite 
constant. One may use as many choice of $\l$ as 
one like. Denote such a set as $\L$.  The set of usable null vectors
is thus given by $\{ \ph^{(\cci)}\} = \{ \ph^{(\ri)}, \ph^{(\l)} \}$
with $\ri =1,2$ and $\l \in \Lambda$.

To proceed, let us 
further partition $(\Ih)$ into
\be
(\Ih) = (\Ih_1, \Ih_2, (\Ih_\l)_{\l\in\L} ), 
\ee
such that $\Ih_\ri = 1, 2, \cdots, N_\ri$,  $\Ih_\l = 1, 2, \cdots, N_\l$
 and $\sum_\ri N_\ri + \sum_\l N_\l =\frN$.
The set of Cartan generators $H_\Ih$ can be relabelled as
\be
\{H_\Ih \} = \{ H_{\Ih_1}, H_{\Ih_2},  \{ H_{\Ih_\l}\}_{\l\in\L}  \}.
\ee
Corresponds to each subset $\{H_{\Ih_\ri} \}$ or $\{ H_{\Ih_\l}\}$, 
one can associate a
set of 1-form roots such that they have nonzero components only when their
indices are in $\Ih_\ri$ or $\Ih_\l$:
\bea \label{1roots-a}
&& \ah^{(\ri)}_I = (0, \cdots,0, \ah^{(\ri)}_{\Ih_\ri},0, 
\cdots, 0;\, 0,0,0,0), \quad \ri=1, 2,\\
&& \ah^{(\l)}_I = (0, \cdots,0, \ah^{(\l)}_{\Ih_\l},0, \cdots, 0;\,
0,0,0,0), \quad \l \in \Lambda \label{1roots-a2}
\eea 
The last four entries refer to the ``internal'' indices $a=1,2,3,4$. 
The 1-forms \eq{1roots-a}, \eq{1roots-a2} 
satisfy the conditions \eq{cond6p} and \eq{cond7p} by
construction and lead to the  following set of two-form roots: 
\bea
&& \a^{(\ri)}= \ph^{(\ri)} \wedge \ah^{(\ri)}, \quad \ri =1,2,  \label{2roots-a}\\
&& \a^{(\l)}= \ph^{(\l)} \wedge \ah^{(\l)}, \quad \l \in \L. \label{2roots-b}
\eea
Or, in terms of the nonzero components, we have 
\bea
&& \a^{(\ri)}_{a \Ih_\ri} = \ph^{(\ri)}_a \ah^{(\ri)}_{\Ih_\ri}\;, \quad\ri =1,2,\\
&& \a^{(\l)}_{a \Ih_\l} = \ph^{(\l)}_a \ah^{(\l)}_{\Ih_\l},  \quad \l \in \L.
\eea

In addition to these roots which have a ``mixed'' indices structure, 
one can also consider  roots with ``internal'' indices. In particular to
satisfy \eq{cond6p} and \eq{cond7p}, the only possibility is have a
root which is a linear combination of $\ph^{(1)}$ and $\ph^{(2)}$. Wedging
it with \eq{p1s} leads to the 2-form roots
\be \label{2roots-r}
r^{(n)} = c_n \ph^{(1)} \wedge \ph^{(2)},
\ee
where $c_n$ is a constant and $n$ is taken from an arbitrary set $X$.
 It is easy to show that
$c(r^{(n)}, r^{(m)}) =0$ for all $n,m \in X$. 

All in all, we obtain the roots \eq{2roots-a}, \eq{2roots-b} 
and \eq{2roots-r}. This
give rises to the step generators
\be
E^{\a^{(\ri)}}, E^{\a^{(\l)}}, E^{\pm r^{(n)}}.
\ee

Next let us regroup the generators of the   
Cartan-Weyl 3-algebra in a way which
will be convenient for our analysis. Let us start with the   
set of Cartan generators 
\be
\{H_I\} = \{ H_{\Ih_1}, H_{\Ih_2},  \{ H_{\Ih_\l}\}_{\l\in\L} , H_a  \}.
\ee
To express the 3-brackets, it is convenient to use 
the following 4 generators instead of the $H_a$'s:
\be
v^{(\ri)} := \ph^{(\ri)} \cdot H, \quad 
u^{(\ri)} := \pt^{(\ri)} \cdot H,\qquad \ri =1,2.
\ee
It is $\la v^{(\ri)},  v^{(\rj)} \ra =0$,  
$\la u^{(\ri)},  u^{(\rj)} \ra =0$,  $\la v^{(\ri)},  u^{(\rj)} \ra
=\d_{\ri \rj}$.
Next we denote the generators
\be
 x_\pm^{(n)}:= E^{\pm r^{(n)}}, 
\ee
where the metric is $\la x_+^{(n)}, x_-^{(m)} \ra = \d_{nm}$. 
It is also convenient to group the step generators and the Cartan
generators in the following manners and introduce the vector spaces
\be
\mg^{(\ri)} := \CC(H_{\Ih_\ri}, E^{\ah^{(\ri)}} ), \quad 
\mg^{(\l )} := \CC(H_{\Ih_\l }, E^{\ah^{(\l )}} ).
\ee
The  Cartan-Weyl 3-algebra of index 2 is thus given by
\be \label{vec2}
\cA= \bigg(\bigoplus_{\ri=1}^2 \mg^{(\ri)} \bigg) \oplus
\bigg( \bigoplus_{\l \in \L} \mg^{(\l)} \bigg) \oplus 
\CC(u^{(1)}, v^{(1)}) \oplus \CC(u^{(2)}, v^{(2)})
\oplus E
\ee
as vector space. Here $E$ is an even dimensional vector space spanned
by the elements $x_\pm^{(n)}$, $n \in X$. 
Different components of the direct sum
in \eq{vec2} are orthogonal to each other. The presence of the vector
space $E$ is a new feature when the index is higher than 1.

\noindent\underline{3-brackets}

Now let us express the   
3-algebra relations \eq{cw31}-\eq{cw34c} in terms of these generators.
Using our results \eq{cond5p} and the result that $c^{(\cci)}$ are
those coefficients for a semisimple Lie algebra, we obtain immediately
the following nonvanishing 3-brackets:
\be \label{fr0}
[u^{(\ri)}, g_1, g_2] = \begin{cases}
[g_1,g_2]_{\mg^{(\ri)}}, & g_1, g_2 \in \mg^{(\ri)},\\
\l_\ri [g_1,g_2]_{\mg^{(\l)}}, &  g_1, g_2 \in \mg^{(\l)},
\end{cases}
\ee
where $\l_1 =1$ and $\l_2= \l$, and 
\be \label{fr1}
[g_1,g_2,g_3] = \begin{cases}
- \la [g_1,g_2],g_3 \ra_{\mg^{(\ri)}} \; v^{(\ri)}, 
& g_1, g_2, g_3 \in \mg^{(\ri)},\\
- \la [g_1,g_2],g_3 \ra_{\mg^{(\l)}} \; (v^{(1)} + \l v^{(2)}), 
& g_1, g_2, g_3 \in \mg^{(\l)}.
\end{cases}
\ee
Also we have
\bea
&& [u^{(1)}, u^{(2)},  x_+^{(n)}] = c_n  x_+^{(n)}, \quad
 [u^{(1)}, u^{(2)},  x_-^{(n)}] = -c_n  x_-^{(n)},  \label{fr2}\\
&& [u^{(1)}, x_+^{(n)}, x_-^{(n)}] = c_n v^{(2)},\quad  
[u^{(2)}, x_+^{(n)}, x_-^{(n)}] = -c_n v^{(1)}. \label{fr2a}
\eea

One may also rewrite \eq{fr2} and \eq{fr2a} in terms of a different basis.
Introducing $x_\pm^{(n)}= (x^{(n)}  \pm i y^{(n)} )/\sqrt{2}$. 
The metric is 
$\la x^{(n)}, x^{(m)} \ra = \la y^{(n)},  y^{(m)} \ra = \d_{nm}$, 
$\la x^{(n)}, y^{(m)} \ra =0$. 
The relations \eq{fr2} and \eq{fr2a} can be 
rewritten in the form
\bea \label{fr3}
&& [u^{(1)}, u^{(2)},x] = Jx, \nn \\
 && [u^{(1)},x,y] = \la Jx,y\ra v^{(2)},\\
 && [u^{(2)},x,y] = -\la Jx,y\ra v^{(1)}, \nn
\eea
where $x,y \in E$ in general and $J$ is an $so(E)$ matrix such that $J
x^{(n)} = i c_n y^{(n)}, J y^{(n)} = -i c_n x^{(n)}$. 

Quite amazingly, the relations 
\eq{fr0}, \eq{fr1} and \eq{fr3} are precisely the same as the
3-algebras obtained in \cite{Jind2} for the index 2 case. There the
3-algebra was constructed by requiring it to have a maximally isotropic center. 
Here we have obtained the same Lie 3-algebra from a different requirement as 
a Cartan-Weyl 3-algebra.

Finally we remark that in addition to the way \eq{m2} in splitting the
metric, we can also split it differently 
\be 
g_{I J} = 
\begin{pmat} ( {..|.} )
 &   &  & &	\cr 
 & g_{\It \Jt} &   &  & \cr
 &   & &  & 		\cr \-
 &   &   & 1& 		\cr
 &   &   &  & -1        \cr
\end{pmat},
\ee
such that  a (1,1) subspace is singled out separately and the metric
$g_{\It \Jt}$ is Lorentzian. Associated with the (1,1) subspace is a
null vector which one can use to construct the two-form roots. 
The construction is exactly the same as for the
the index 1 case except that $g_{\It \Jt}$ is now non-Euclidean.  
One obtain immediately the Lorentzian 3-algebra $\cA= \mg \oplus \CC(
u,v)$ 
with $\mg$ being a semisimple Lie algebra of index 1. This is nothing
new. 

\subsection{Higher index $m \geq 3$}

The above construction can be generalized to  index 3 or higher
easily. First let us choose a basis of the Cartan generators
such that the metric takes the form with nonzero entries
\be
g_{I J} = 
\begin{pmat} ( {|.} )
g_{\Ih \Jh} &     \cr \- 
 & l_{ab}  \cr
\end{pmat},
\ee
where $g_{\Ih \Jh}$ is
Euclidean and $l_{ab}$ is a 2$m$-dimensional 
metric with signature $(m,m)$. The set of Cartan
generators is $\{ H_I \} = \{ H_\Ih ; H_a\}$ with $ \Ih =1, \cdots, \frN$;  
$a, =1,
\cdots, 2m$. The rank is $N=\frN+2m$. 

\noindent \underline{root system}

As before, to solve \eq{cond6p}, we can 
take the 1-form roots $\a_I =(\a_\Ih; {\bf 0} )$ 
and the null vectors in the orthogonal
space: $\ph =({\bf 0}; \cdot)$.  There are $2m$ such linearly independent
null vectors. One can choose a basis of null
vectors $\{ \ph^{(\ri)}, \pt^{(\ri)}\}$, $\ri =1, \cdots, m$ such that 
\be
\ph^{(\ri)} \cdot \ph^{(\rj)} = 0, \quad
\pt^{(\ri)} \cdot \pt^{(\rj)} = 0, \quad
\ph^{(\ri)} \cdot \pt^{(\rj)} = \d_{\ri\rj}, \qquad
\ri, \rj  =1, \cdots, m.
\ee
In addition to the $\ph^{(\ri)}$'s, one can also solve 
\eq{cond6a} and \eq{cond6} with the following null vectors
\be
\ph^{(\l)} := \sum_\ri \l_\ri \ph^{(\ri)},
\ee
where $\l_\ri$'s are arbitrary constants. 
One can use as many choice of $(\l_\ri)$ as one like. Denotes such a set as
$\Lambda$. The set of usable null vectors
is given by $\{ \ph^{(\cci)}\} = \{ \ph^{(\ri)}, \ph^{(\l)} \}$
with $\ri =1,\cdots, m$ and $(\l_\ri) \in \Lambda$.

Using these null vectors, one can construct solution to \eq{cond6p} and
\eq{cond7p} by  first partitioning $(\Ih)$ into many parts
\be
(\Ih) = ((\Ih_\ri)_{\ri =1, \cdots, m} \; , (\Ih_\l)_{\l \in \Lambda} ), 
\ee
such that $\Ih_\ri = 1, 2, \cdots, N_\ri$ for $\ri=1, \cdots,m$,  
 $\Ih_\l= 1, 2, \cdots, N_\l$ and $\sum_\ri N_\ri +\sum_\l N_\l =\frN$.
The set of Cartan generators $H_\Ih$ can be relabelled as
\be
\{H_\Ih \} = \{ \{H_{\Ih_\ri}\}_{\ri =1, \cdots, m} \;, 
\{H_{\Ih_\l}\}_{\l \in \Lambda}  \}.
\ee
Corresponds to each subset $\{H_{\Ih_\ri} \}$, one can associate a
set of 1-form roots such that they have nonzero components only when their
indices are in $\Ih_\ri$ or $\Ih_\l$
\bea 
&& \ah^{(\ri)}_I = (0, \cdots,0, \ah^{(\ri)}_{\Ih_\ri},0, \cdots, 0;\,
{\bf 0}),    \label{1roots-ap1} \\
&& \ah^{(\l)}_I = (0, \cdots,0, \ah^{(\l)}_{\Ih_\l},0, \cdots, 0;\,
{\bf 0}). \label{1roots-ap2}
\eea 
The last entries refers to the ``internal'' indices $a=1, \cdots, 2m$. 
The set \eq{1roots-ap1}, \eq{1roots-ap2} 
lead to the  following set of two-form roots: 
\bea 
\a^{(\ri)}&=&  \ph^{(\ri)} \wedge \ah^{(\ri)}, \quad \ri =1
,\cdots, m,\label{2roots-ap1} \\
\a^{(\l)}&=& \ph^{(\l)} \wedge \ah^{(\l)}, \quad (\l) \in \Lambda .
\label{2roots-ap2} 
\eea

In addition to these roots which have a ``mixed'' indices structure,
one can also consider  roots $\a_{ab}$ with only ``internal'' indices. 
In particular to
satisfy \eq{cond6p} and \eq{cond7p}, the only possibility is to use 1-form
roots which are linear combination of $\ph^{(\ri)}$. Wedging
it with $\ph^{(\ri)}$ or $\ph^{(\l)}$ leads to the 2-form roots
of the form 
\be \label{2roots-rp}
r^{(\mu)} = \sum_{\ri,\rj} \mu_{\ri \rj}\;  \ph^{(\ri)} \wedge \ph^{(\rj)},
\ee
where $\m_{\ri \rj}$ are  constants and not all equal to zero. There
is an internal root \eq{2roots-rp} corresponds to 
each  choice of $(\mu_{\ri \rj})$. Denote the set of $(\m)$ by $X$.
It easy to see that one should take
$c(r^{(\m)}, r^{(\m')}) =0$  for all $(\m),(\m') \in X$.

The roots  \eq{2roots-ap1}, \eq{2roots-ap2} and \eq{2roots-rp} 
give rises to the step generators
\be
E^{\a^{(\ri)}},  E^{\a^{(\l)}}, E^{\pm r^{(\mu)}}.
\ee

As before, let us regroup the generators of the  
3-algebra in a way which
will be convenient for our analysis. Let us start with the   
set of Cartan generators. It is convenient to use 
the following $2m$ generators instead of the $H_a$'s:
\be
v^{(\ri)} := \ph^{(\ri)} \cdot H, \quad 
u^{(\ri)} := \pt^{(\ri)} \cdot H,\qquad \ri =1, \cdots, m.
\ee
It is $\la v^{(\ri)},  v^{(\rj)} \ra =0$,  
$\la u^{(\ri)},  u^{(\rj)} \ra =0$,  $\la v^{(\ri)},  u^{(\rj)} \ra
=\d_{\ri \rj}$.
Next we introduce the notations
\be
 x_\pm^{(\m)}:= E^{\pm r^{(\m)}}, 
\ee
where the metric is $\la x_+^{(\m)},  x_-^{(\m')} \ra = \d_{\m \m'}$. 
It is also convenient to group the step generators and the Cartan
generators in the following manners and introduce the vector spaces
\be
\mg^{(\ri)} := \CC (H_{\Ih_\ri}, E^{\ah^{(\ri)}} ), \quad
\mg^{(\l)} := \CC (H_{\Ih_\l}, E^{\ah^{(\l)}} ).
\ee
The  Cartan-Weyl 3-algebra of index $m \geq 3$ 
is thus given by
\be \label{vec2p}
\cA= \bigg(\bigoplus_{\ri=1}^m \mg^{(\ri)} \bigg) \oplus
\bigg(\bigoplus_{\l \in \Lambda} \mg^{(\l)} \bigg) \oplus
\bigg( \bigoplus_{\ri=1}^m \CC(u^{(\ri)}, v^{(\ri)}) \bigg) \oplus 
E
\ee
as vector space. Here $E$ is a even dimensional vector space spanned
by the elements $x_\pm^{(\m)}, (\mu) \in X$. 
Different components of the direct sum
in \eq{vec2p} are orthogonal to each other.

\noindent\underline{3-brackets}

The 3-brackets can be worked out similarly and we obtain the following
nonvanishing relations for the  Cartan-Weyl 3-algebra:
\be \label{gen31}
[u^{(\ri)}, g_1, g_2] = \begin{cases}
[g_1,g_2]_{\mg^{(\ri)}}, & g_1, g_2 \in \mg^{(\ri)},\\
\l_\ri [g_1,g_2]_{\mg^{(\l)}}, &  g_1, g_2 \in \mg^{(\l)},
\end{cases}
\ee
\be\label{gen32}
[g_1,g_2,g_3] = \begin{cases}
- \la [g_1,g_2],g_3 \ra_{\mg^{(\ri)}} \; v^{(\ri)}, 
& g_1, g_2, g_3 \in \mg^{(\ri)},\\
- \la [g_1,g_2],g_3 \ra_{\mg^{(\l)}} \; (\sum_\ri \l_\ri v^{(\ri)}), 
& g_1, g_2, g_3 \in \mg^{(\l)},
\end{cases}
\ee
\bea 
[u^{(\ri)}, u^{(\rj)},  x_\pm^{(\m)}] 
&=& \pm \m_{\ri \rj} x_\pm^{(\m)}, \label{gen33} \\
\; [u^{(\ri)}, x_+^{(\m)},  x_-^{(\m)}] &=& \sum_\rj \m_{\ri \rj}  v^{(\rj)}.
 \label{gen34}
\eea

It is quite remarkable that, starting out as a natural generalization
of a Cartan-Weyl algebras, 
Cartan-Weyl 3-algebras as defined by 
\eq{cw31}-\eq{cw34c} can be  constructed and classified 
completely.

\section{Remarks on  Fuzzy $S^3$ in BLG Theory}

In the original construction of BLG, 
the Lie 3-algebra  $\cA_4$ was employed.
The metric on $\cA_4$ is positive definite and the BLG
model defines a unitary quantum field theory. 
The use of 
$\cA_4$  was  motivated by the studies of 
Basu and Harvey \cite{BH} whose main objective was to construct a
generalization of the Nahm equation for describing intersecting M-branes.
Employing a Lie 3-algebraic structure of $\cA_4$, the Basu-Harvey 
equation  admits a solution whose cross section is given by a fuzzy
$S^3$ and describes
the puffing up of a system of multiple M2-branes into a M5-brane.

In general, one would like to employ more general metric Lie 3-algebras 
with arbitrary higher ranks in the 
BLG description of multiple M2-branes\footnote{
If one is interested only in the supersymmetric equations
of motion, it is possible to do without a metric \cite{nilsson}.}.
The choice of the
Lie 3-algebras should be such that the BLG theory or 
generalization of the original construction, gives  a  unitary Quantum 
Field Theory. The Lie 3-algebra should also allows different embedding of
$\cA_4$ in it in much the same way as one can find 
fuzzy $S^2$  of different sizes in $SU(N)$. 
Presumably the different $\cA_4$'s would then be 
characterized by some kinds of Casimir of the $\cA_4$ algebra 
which corresponds to 
having different number of M2-branes puffing up to a single M5-brane 
in the Basu-Harvey's fuzzy funnel solution of the multiple
M2-branes system. 
However achieving these objectives turns out to be highly nontrivial.
For discussions about unitary BLG theory, see 
\cite{LOR,GP,GG,Jind2,Jindn,d2d2,HMS}.  
At present,  manifest unitary remains a major obstacle of the BLG theory. 
To resolve it will require the use of a different kind of metric 
Lie 3-algebras together with a novel ghost decoupling mechanism.
This is beyond the scope of this work and we will have nothing 
more to say on this problem.

The problem of finding a fuzzy $S^3$ solution 
has been considered recently for the ABJM theory \cite{ram}. 
Surprisingly to the best
of our knowledge, the question of finding a fuzzy $S^3$ 
for the BLG theory in general has not been 
considered before. We will examine  this issue 
in details now. 

It is
instructive to first recall the case of multiple D1-branes. There
the Nahm equation, the BPS
equation for the $U(N)$ non-abelian Born-Infeld theory of the
D1-strings, admits a fuzzy funnel whose transverse 
cross section is a fuzzy $S^2$ and describes a bunch of D1-strings 
puffing up into a D3-brane \cite{myers}, corresponding to a D1-D3 brane
intersection in string theory. Similarly, one would like to be able to
describe the M2-M5 branes
intersection as a fuzzy funnel solution of a BPS equation of 
the multiple M2-branes theory.  In fact this was the original
motivation leading Basu and Harvey \cite{BH} to write down such a BPS
equation. It has been demonstrated in \cite{BL3} that the BLG
Lagrangian based on the algebra $\cA_4$ admits a fuzzy funnel solution whose
energy density scales as expected of a M2-M5 intersection (both 
the massless  and massive cases). 
However the problem has not been discussed for 
BLG theory based on more general Lie 3-algebras, e.g. the
Lorentzian 3-algebra.

In general  a
fuzzy $S^3$ solution in the BLG theory 
is described by having $X^\cP$, $\cP=1,2,3,4$ satisfying the $\cA_4$
algebra
\be \label{A4}
[X^\cP, X^\cQ, X^\cR] = i \e^{\cP \cQ \cR}{}_\cS X^\cS. 
\ee
This describes a $SO(4)$ invariant distribution in the theory. In addition
one needs also a certain ``size'' condition so that one can be sure one is
describing a single fuzzy $S^3$. In the case of fuzzy $S^2$ for
D1-D3 intersection, the size condition is written down with respect
to the representation taken by the $X^\cP$'s which specifies the radius
of the fuzzy $S^2$. For fuzzy $S^3$, one can expect that one will 
need to generalize the concept of a representation of Lie 3-algebras in order to
specify a suitable ``size'' condition. It is not known how to write
down arbitrary representations for a  general Lie 3-algebra. See \cite{wolf} 
for some discussions.
However it is possible that for certain specific kinds of Lie 3-algebras,
for example, such as the Cartan-Weyl 3-algebras due to the Lie algebraic
structure they inherited.
In fact,
because of \eq{metric-cA} and \eq{metric3-eeh}, the invariant 
metric of the Cartan-Weyl
3-algebra is precisely the same as the invariant metric 
of the underlying Lie algebra. 
Since the later can be immediately generalized to the Killing metric
which is defined for an arbitrary
representation of the Lie algebra, this can be extended to
the Cartan-Weyl 3-algebras immediately. 
In the following, however, we will focus on the algebraic condition \eq{A4}
only  which is
independent of any representation issues.

Let us first  consider  the Lorentzian 3-algebra \eq{i1-5}. Let $X^\cP=
X^\cP{}_u u + X^\cP{}_v v+ X^\cP{}_\ah g^\ah $, where the Lie algebra
generators $g^\ah$ obeys $[g^\ah, g^\bh] = f^{\ah\bh}{}_\gh g^\gh$.
The equation \eq{A4} gives
\bea
X^\cP{}_u &=& 0, \nn\\
 i \e^{\cP \cQ \cR}{}_\cS X^\cS{}_v &=& - X^\cP{}_\ah X^\cQ{}_\bh X^\cR{}_\gh 
f^{\ah\bh\gh} ,
\\  
 i
\e^{\cP \cQ \cR}{}_\cS X^\cS{}_\dh  &=&
X^\cP{}_\ah X^\cQ{}_\bh X^\cR{}_u f^{\ah\bh}{}_\dh 
+ \mbox{($\cP,\cQ,\cR$ cyclic)}.\nn
\eea  
Hence we obtain $X^\cS{}_u = X^\cS{}_\dh =X^\cS{}_v =0$ and 
the Lorentzian BLG theory does not admit any fuzzy $S^3$ solution.

The above consideration can be generalized immediately to the general
Cartan-Weyl 3-algebra \eq{cw31}-\eq{cw34c}.  
Let $X^\cP= X^\cP{}{}^I H_I + X^\cP{}_\a E^\a $.  
The equation \eq{A4} gives
\bea  
i \e^{\cP \cQ \cR}{}_\cS X^\cS{}^L g_{LI} &=& 
-\sum_{\a+\b+\g=0} X^\cP{}_\a X^\cQ{}_\b X^\cR{}_\g g_I(\a,\b) \nn\\
&&\quad + \sum_{\b} X^\cP{}^L X^\cQ{}_\b X^\cR{}_{-\b} \b_{LI}  
+\mbox{($\cP,\cQ,\cR$
  cyclic)}, \label{XXX1} \\
i \e^{\cP \cQ \cR}{}_\cS X^\cS{}_\a &=& X^\cP{}^I X^\cQ{}^J X^\cR_\a \a_{IJ} \nn\\  
&& \quad + \sum_\b X^\cP{}^I X^\cQ{}_\b X^\cR_{\a-\b} g_I(\b, \a-\b) 
+\mbox{($\cP,\cQ,\cR$
  cyclic)}.\;\;\;\;\;\; \label{XXX2}
\eea
Now given the general solution \eq{cond-ap}, \eq{cond-gI}, we can
multiply the equation \eq{XXX1} with $\ph^{(\cci)}{}^I$ and using the
properties \eq{cond6ap}, \eq{cond6p} to obtain that
\be
X^\cS{}^I \ph^{(\cci)}{}_I=0, \quad \mbox{for all $\cci$ and $\cS$}.
\ee
Substituting this into \eq{XXX2}, we than obtain  $X^\cS{}_\a
=0$. This, substituting back into \eq{XXX1}, then implies that $X^\cS{}^I  =0$. 
Therefore the Cartan-Weyl 3-algebra 
does not admit  $\cA_4$ as a subalgebra. 
We remark that this is quite different from the case of  Lie algebra where
any Lie algebra admits $SU(2)$ as a subalgebra.

The problem of finding a fuzzy $S^3$ solution
has also been considered in \cite{ram} for the ABJM
theory. For large value of the level $k$  where one can
trust the semi-classical analysis, it was found that 
the ABJM theory admits only a fuzzy $S^2$ structure, rather than a
fuzzy $S^3$. This is the expected result since for generic $k$, the
geometry is $S^3/Z_k$ and the M-theory circle $S^1/Z_k$ becomes
zero in the $k \to \infty$ limit, reducing the system to the D2-D4
branes intersection in IIA string theory. Since the large $k$ limit is
needed in order to have a perturbative formulation of the ABJM theory, 
finding a  fuzzy $S^3$ as the (semi-)classical geometry in the ABJM
theory is impossible. 

For the BLG theory, there is a similar
problem. Presumably the level $k$  corresponds to some order of
orbifolding \cite{orbifold} 
and a theory of multiple M2-branes in flat space is given by a 
level one BLG theory with a certain choice of Lie 3-algebras. 
It is not clear whether the fuzzy $S^3$ should emerge as  a classical
solution as in the original BLG theory \cite{BL3} or only  as a solution 
of the full quantum system as in the ABJM theory. 
Since the former situation is what happened in the original BLG
theory, it might be natural to expect that this to be also the case for 
the BLG theories based on more general Lie 3-algebras. 
Assuming that this is the case,
this no-go theorem 
of finding a fuzzy $S^3$ solution in the Cartan-Weyl 3-algebras 
then means that one need to consider more general class 
of Lie 3-algebras for use 
in the BLG theory. In the companion paper \cite{cat2}, 
a certain generalization of the Cartan-Weyl 
3-algebras is suggested and we will show that 
the no-go theorem can be bypassed easily in this class of Lie 3-algebras, giving 
the hope of having fuzzy $S^3$ solution.

\section{Discussions}

In this paper we have generalized the notion of a Cartan-Weyl basis for a
Lie $3$-algebra. We have also shown that the consistency conditions 
defining a Cartan-Weyl 3-algebra can be solved exactly, leading to a complete
classification of Cartan-Weyl 3-algebras. This is the main result of this paper.
It is natural to speculate that Cartan-Weyl 3-algebras may be useful 
for describing 
some  kinds of generalized symmetry. It will be
interesting to discover more of this.

We have mostly worked with metric Lie 3-algebras in this paper. 
It is clear that
the concept of a Cartan-Weyl
basis can be similarly defined for  higher metric Lie $n$-algebras and  
the existence of a Cartan-Weyl basis  is
equivalent to the requirement that the solutions of the 
``eigenvalue equation''
\be
[H_{I_1}, \cdots, H_{I_{n-1}}, Y^a] = \psi_{I_1 \cdots  I_{n-1}}{}^a{}_b Y^b
\ee
are non-degenerate, and that the
restriction of the metric to the $H$'s is non-degenerate. As a result, a 
Cartan-Weyl $n$-algebra is equipped with a number of
mutually commuting Cartan generators $H_I$
together with a number of step generators $E^\a$
parametrized by a root space of non-degenerate $(n-1)$-forms $\a$.
As we have seen above, Cartan-Weyl
3-algebras have curiously built in them a structure of semisimple Lie
algebras. One can expect that Cartan-Weyl $n$-algebras will also
have built in them a certain special kind of Lie $(n-1)$-algebras. 
It will be interesting to solve the
corresponding 
consistency conditions and construct the Cartan-Weyl $n$-algebras
explicitly and find this out. 
We speculate that these kinds of Lie $n$-algebras may
have a good chance to be of use in physics.  

In order to understand which kind of Lie 3-algebras (or Lie
$n$-algebras in general) might appear in  physical
descriptions, it will be very helpful to understand how to
``integrate'' the infinitesimal transformations described by 
Lie 3-algebras to 
finite transformations since it is usually 
much more clear how 
the finite transformations should be constrained. One may call these 
``Lie 3-group'' transformations.  
There are many related interesting mathematical questions one may ask.
For example, Lie 3-group is certainly not a Lie group. 
How is a Lie 3-group defined? It seems natural that  Lie
3-bracket may have its origin in a tertiary, perhaps nonassociative, product
structure of a Lie 3-group. Is it true? 
Is there a Baker-Hausdorff formula? 
While fascinating, there is nothing known in the literature 
about how to think about a Lie 3-group.
Therefore we will have to look for other traits in order
to identify the desired type of Lie 3-algebras that could be relevant 
for physical applications.

For applications in the 
BLG models, the prospect of  Cartan-Weyl 3-algebras is not good 
since the corresponding BLG
models are equivalent to ordinary Yang-Mills theories and so, not
surprisingly, they do not admit any fuzzy $S^3$ structure. Such BLG
models describe D-branes rather than multiple M2-branes. 
We will argue in the paper  \cite{cat2} that the Lie 3-algebras of
interest 
should satisfy a certain reduction condition. We will show that
this reduction condition leads to a class of metric Lie 3-algebras which 
naturally generalizes the Cartan-Weyl 3-algebras  introduced 
in this paper. 
These  {\it generalized Cartan-Weyl 3-algebras} 
have the same form \eq{ccw2}
of the Lie 3-brackets. However their Cartan subalgebra can be 
non-abelian in general, i.e. the 3-brackets $[H_I,H_J,H_K] \neq 0$. 
This modification is indeed a welcome one. We will explain in \cite{cat2} 
how this modification may help with 
getting fuzzy $S^3$ solutions  in the  corresponding BLG models.

\section*{Acknowledgements}

It is a pleasure to thank Douglas Smith for many useful discussions and 
comments.  
The research is supported by EPSRC and STFC.

\appendix

\section{Consistency Conditions of Cartan-Weyl 3-algebras} 
\label{gen-const}
 
To construct a  Cartan-Weyl 3-algebra, 
one need to specify the system
of two-form roots $\a_{IJ}$ and the coefficients $g_I(\a,\b)$,
$c(\a,\b,\g)$. These are constrained by the consistency of 
\eq{cw31}-\eq{cw34c}
with the fundamental identity (FI). Let us examine this in details now.

\noindent\underline{$[[H_I, H_J, H_K], H_L, E^\a] =0$:}

Consistency of \eq{cw31} with the FI gives the condition
\be \label{cond1}
\a_{IJ} \a_{KL} + \a_{JK} \a_{IL} + \a_{KI} \a_{JL} =0.
\ee 
This condition is nontrivial only if the rank is at least four. 
For this general case, 
let us introduce auxiliary differentials $dx^K$, then the condition can
be written as
\be
\a \wedge i_\xi \a =0,
\ee
where $\a := \a_{IJ} dx^I dx^J$ is a two form and 
$i_\xi$ is the Cartan inner
product with $\xi$ being an arbitrary vector. 
It is then clear that
the general  solution of \eq{cond1} is given by 
\be \label{rootaa}
\a = \a_1 \wedge \a_2
\ee 
where 
$\a_1, \a_2$ are arbitrary one-forms. In terms of components,
it is $\a_{IJ} = \a_{1I} \a_{2J} - \a_{2I} \a_{1J}$. We remark that
although the condition \eq{cond1} is trivial for the case of $N=3$,
the 2-form root (dual to a vector) 
can still be written in the form of \eq{rootaa}. The same is true for
$N=2$. So we conclude that the 2-form roots of a  
Cartan-Weyl 3-algebra always take the form
\eq{rootaa}.

\noindent\underline{$[[H_I, H_J, H_K], E^\a, E^\b] =0$:}

Consistency with the FI gives the condition
\be \label{cond2p}
(\a+\b)_{JK} g_I(\a,\b) + (\a+\b)_{KI} g_J(\a,\b) + (\a+\b)_{IJ}g_K(\a,\b)  =0. 
\ee 

\noindent\underline{$[[H_I, H_J, E^\a], H_K, H_L] =0$:}

Consistency of \eq{cw32}  with the FI is satisfied without the need of any
new condition.

\noindent\underline{$[[H_I, H_J, E^\a], H_K, E^\b] =0$:}

There are two cases to consider.  For the case 
$\a+\b=0$, consistency with the FI is
satisfied due to \eq{cond1}. For $\a+ \b \neq 0$, we get the new
condition
\be \label{cond2}
\b_{JK} g_I(\a,\b)  + \b_{KI} g_J(\a,\b)  + \b_{IJ} g_K(\a,\b)  =0. 
\ee 
The can be written as  
\be 
\b \wedge g(\a,\b) =0
\ee
with $g:=g_I dx^I$.
By symmetry we also have $\a \wedge g(\a,\b) = 0$ and the condition
\eq{cond2p} is a consequence of \eq{cond2}. Note that for a 
 Cartan-Weyl 3-algebra
with only one linearly independent root i.e. with a pair of roots $\a,
-\a$, the condition \eq{cond2p} is satisfied automatically 
and there is no \eq{cond2}. 
\begin{lem}
Denote by $\Omega$ the set of roots such that $g_I(\a,\b) \neq 0$ for $\a,\b
\in \Omega$. 
The condition \eq{cond2} implies that  in general the roots $\a$ and 
$g_I$ take the following factorized form on $\Omega$
\be
\a= \ph \wedge \hat{\a} \label{cond3}  ,
\ee
\be
 g_I(\a,\b) = \ph_I c(\a,\b)  \label{cond4} ,
\ee
where $\ph_I$ is a fixed one-form 
and $c(\a,\b)$ is a scalar function of the roots.  
\end{lem}
\begin{proof}
Let $\a= \a_1 \wedge \a_2$,  $\b= \b_1 \wedge \b_2$, 
$\g= \g_1 \wedge \g_2$ be any 3 roots such that $g_I$ is nonvanishing
when evaluated on any two of them.
The conditions $\a \wedge g(\a,\b) = \b \wedge g(\a,\b) =0$ with $g:=g_I dx^I$,
is solved by
\be
g(\a,\b) = \a_1 + \m \a_2, \qquad g(\a,\b) = \b_1 + \l \b_2,
\ee
for some numbers $\m, \l$. Some of the coefficients have been chosen to 1 by 
a proper normalization of $\a_1, \a_2$ etc. 
Similarly since $g(\a,\g) \neq 0$, then we
obtain
\be
g(\a,\g) = \m'\a_1 + \m'' \a_2, \qquad g(\a,\g) = \g_1 + \l' \g_2.
\ee
And since $g(\b,\g) \neq 0$,  we have
\be 
g(\b,\g) = c_1 \b_1 + c_2 \b_2, \qquad g(\b,\g) = c_3 \g_1 + c_4 \g_2.
\ee
It follows immediately from the consistency of these equations that
$g(\a,\b)$, $g(\b,\g)$, $g(\a,\g)$ are all proportional to each other. Hence
for  any four roots $\a,\b,\g,\d$, we have $g(\a,\b) \propto  g(\g,\d)$
provided that $g(\cdot,\cdot)$ is not equal to zero when evaluated on
any two of the four roots. This can be satisfied only if \eq{cond4}
holds. Finally the condition $\a\wedge g =0$ implies \eq{cond3}.
\end{proof}
Note that here the choice of $\ph$ and $c$ depends on $\Omega$. Most generally, 
the set $\D_\cA$ of all roots can be divided
into a number of disjoint subsets $\Omega_\cci$ 
characterized by an orthogonality condition:  $g(\a, \b) =0$ 
if $\a \in \Omega_\cci$ and
$\b \in \Omega_\ccj$ with $\cci \neq \ccj$. In this
case, we have
\be \label{cond5}
\boxed{
\a = \ph^{(\cci)} \wedge \ah^{(\cci)}, \quad \a \in \Omega_\cci, 
}\  
\ee
and
\be \label{cond5p}
\boxed{
g_I (\a,\b)  = \begin{cases}
\ph_I^{(\cci)} c^{(\cci)}(\ah^{(\cci)},\bh^{(\cci)}), & \quad \mbox{for $\a,\b \in
\Omega_\cci$}, \\
0, & \quad \mbox{otherwise}.
\end{cases}
}
\ee
Thus associated with each orthogonal component $\Omega_\cci$, there is a
function $c^{(\cci)}$ and a fixed one form $\ph^{(\cci)}$.

\noindent\underline{$[[H_I, H_J, E^\a], E^\b, E^\g] =0$:}

There are three cases to consider. 
The case where $\a+\b+\g =0$ does not give rise to any new condition.
The case  $\a+\b+\g \neq 0$ and $\b+ \g \neq 0$ gives the condition
\be
c(\a,\b,\g) =0. 
\ee
The case  $\a+\b+\g \neq 0$ and $\b+ \g = 0$ gives the condition
\be \label{e1}
\a_{IL} \b_{JK} g^{KL} = \a_{JL} \b_{IK} g^{KL} \quad \mbox{for all $\a,\b$}.
\ee
More explicitly, for $\a = \ph^{(\cci)} \wedge \ah^{(\cci)}$,
$\b=\ph^{(\ccj)} \wedge \bh^{(\ccj)}$, this  condition reads
\be \label{cond6a}
\boxed{
\ph^{(\cci)} \cdot \ph^{(\ccj)} =0, 
\quad \mbox{for all $\cci,\ccj$},
}
\ee
\be \label{cond6}
\boxed{\ph^{(\cci)} \cdot \ah^{(\ccj)} =0,
\quad \mbox{for all $\cci,\ccj$},
} 
\ee
\be\label{cond7} \boxed{
\ah^{(\cci)} \cdot \bh^{(\ccj)} =0, \quad \mbox{for $\cci \neq \ccj$},
}
\ee
where the dot product is taken with respect to the metric $g_{IJ}$. 
As consequences, it is easy to deduce that
\bea
\a_{IK} g^{KL}g_L(\g,\d) &=& 0,  \label{e2}\\
g_K(\a,\b) g^{KL} g_L(\g,\d) &=&0, \label{e3}
\eea
for all roots $\a, \b, \g, \d$. We note that, again, if the algebra has only
a single pair of roots $\pm \a$, then \eq{e1} is trivially satisfied and
\eq{cond6}, \eq{cond7} are not needed. 

\noindent\underline{$[[H_I,E^\a, E^\b], H_K, H_L] =0$ (with $\a+\b =0$):}

Consistency of \eq{cw33a} with the FI is satisfied without
the need of any new condition.

\noindent\underline{$[[H_I,E^\a, E^\b], H_J, E^\g] =0$ (with $\a+\b =0$):}
  
There are two cases to consider. For the case where 
$\a+\g =0$ and $\b+\g \neq 0$, consistency is satisfied automatically.
For the case where $\a+\g \neq 0$ and $\b+\g \neq 0$, we obtain the
condition
\be  
c(\a,\g) c(-\a,\g+\a) - c(-\a, \g) c(\a,\g -\a) = -  \ah \cdot \gh .
\ee
When $\a$ and $\g$ belongs to different components $\Omega_i$ of the 
root space, this equation is trivially satisfied if 
\be
\ah^{(\cci)} \cdot \gh^{(\ccj)} =0, \quad \mbox{if $\cci \neq \ccj$}.
\ee
When they belongs to
the same component of the root space, the condition reads
\be \label{cond-cc1}
\boxed{
c(\ah,\gh) c(-\ah,\gh+\ah) - c(-\ah, \gh) c(\ah,\gh -\ah) = - \ah \cdot
\gh .
}
\ee  
Again this condition does not apply if there is only one linearly
independent root. 

\noindent\underline{$[[H_I,E^\a, E^\b], E^\g, E^\d] =0$  (with $\a+\b =0$):}

There are two cases to consider. For the case where $\g+\d =0$,
consistency with FI is satisfied due to \eq{e1}. For the case where
$\g+\d \neq 0$, consistency with FI is satisfied due to \eq{e2}. 

\noindent\underline{$[[H_I,E^\a, E^\b], H_J, H_K] =0$ (with $\a+\b
  \neq 0$):}

Consistency with FI is satisfied without the need of any new condition.

\noindent\underline{$[[H_I,E^\a, E^\b], H_J, E^\g] =0$ (with $\a+\b
  \neq 0$):}

There are three cases to consider.
First the case $\a+\b+\g =0$. Since $\a+\g \neq 0$ 
and $\b+\g \neq 0$, 
one can easily see that
consistency with the FI is satisfied due to \eq{cond2}. Next consider 
$\a+\b+\g  \neq 0$, we have two subcases. In the first case where $\a+
\g =0$ and $\b+\g \neq 0$, we require the condition \eq{cond-cc1}.
For the second case where $\a+ \g \neq 0$ and $\b+\g  \neq 0$, we obtain
the condition 
\be
g_I(\g,\b) g_J(\g,\a) = g_I(\g,\a) g_J(\g,\b) .
\ee
This is satisfied due to \eq{cond4}.

\noindent\underline{$[[H_I,E^\a, E^\b], E^\g, E^\d] =0$ (with $\a+\b
  \neq 0$):}

One can check that  FI is satisfied without the need of any new
condition.

\noindent\underline{$[[H_I,E^\a, E^\b], \cdot ,\cdot   ] =0$ 
($\a+\b$ not a root):}

Similarly one can check that FI is satisfied without the need of any new
condition.

\noindent\underline{$[[E^\a,E^\b, E^\g], \cdot, \cdot ] =0$ 
(with $\a+\b+ \g =0$):}

One can check that  FI is satisfied without the need of any new
condition.

\noindent\underline{$[[E^\a,E^\b, E^\g], H_I, H_J] =0$
(with $\a+\b+ \g \neq 0$):}

One can check that  FI is satisfied without the need of any new
condition.

\noindent\underline{$[[E^\a,E^\b, E^\g], E^\d, E^\e] =0$ 
(with $\a+\b+ \g \neq 0$):}

There is only one nontrivial case to consider, that is when
$\a+ \d+\e \neq 0$, $\b+ \d+\e \neq 0$ and $\g+ \d+\e \neq 0$. In this
case, it is easy to check that FI is satisfied due to the condition \eq{e3}.

\noindent\underline{$[[E^\a,E^\b, E^\g], H_K, E^\b] =0$ 
(with $\a+\b+ \g \neq 0$):}
 
The check here is more involved. One can check that the FI is satisfied
for most of the combination of $\a,\b,\g$ and $\d$ except for two
cases:\\
(i) when $\a+\d=0$, $\b+ \g =0$, $\b+\d \neq 0$ and $\g +\d \neq 0$, one
needs the condition
\be \label{cond-cc2p}
c(\a,\b) c(-\a, -\b) - c(\a, -\b) c(-\a, \b) = \ah \cdot \bh .
\ee
The condition is trivially satisfied when $\a, \b$ belong to 
different components 
of the root space. When they belong to the same component, the
condition reads
\be \label{cond-cc2}
\boxed{
c(\ah,\bh) c(-\ah, -\bh) - c(\ah, -\bh) c(-\ah, \bh) = \ah \cdot \bh . 
}
\ee
(ii) when $\a+\d= -\b-  \g  \neq 0$, $\b+\d \neq 0$ and $\g +\d \neq 0$, one
needs the condition
\be \label{cond-cc3}
c(\a, -\a-\b-\g) c(\b,\g) + c(\b,-\a-\b-\g) c(\g,\a) +
c(\g,-\a-\b-\g)c(\a,\b) =0 . 
\ee
Note that since 
\be \label{ccc}
c(\a,\b) = c(\b,\g) =c(\g,\a) \quad \mbox{for $\a+\b+\g =0$},
\ee 
the condition \eq{cond-cc3} can be rewritten as 
\be 
c(\a,\b) c(\g, \a+\b) + c(\b,\g) c(\a,\b+\g) + c(\g,\a) c(\b,\g+\a) =0.
\ee
Again, the condition is trivially satisfied when $\a, \b, \g$ 
belong to different components of the root space. 
When they belong to the same component, the condition reads
\be \label{cond-cc4}
\boxed{
c(\ah,\bh) c(\gh, \ah+\bh) + c(\bh,\gh) c(\ah,\bh+\gh) + c(\gh,\ah)
c(\bh,\gh+\ah) =0.
}
\ee

\noindent\underline{$[[E^\a,E^\b, E^\g], \cdot, \cdot] =0$
($\a+\b+ \g$ not a root):}

One can check that  FI is satisfied without the need of any new
condition.

Now we proceed to solve the conditions \eq{cond-cc1}, \eq{cond-cc2} 
and \eq{cond-cc4}. 
We first note that  \eq{cond-cc2} is indeed 
equivalent to \eq{cond-cc1} due to \eq{ccc}.  
Next we note that the condition \eq{cond-cc1} is indeed 
precisely the same condition as one would impose for a semisimple Lie algebra. 
To see this, consider a semisimple Lie algebra with generators 
$\{E^\ah, H_I \}$, metric
\be \label{il1-2}
\la H_I, H_J \ra :=g_{IJ}, \quad 
\la E^\ah, E^\bh \ra =  \d_{\a+\bh},\quad
\la E^\ah, H_I \ra =0,
\ee
and the Lie brackets
\bea \label{il1-1}
\; && [H_I,H_J] =0, \nn\\
\; && [H_I,E^\ah] = \ah_I E^\ah, \\
\; && [E^\ah,E^\bh] = \begin{cases}
\ah\cdot H & \mbox{for $\ah+\bh =0$},\\
c(\ah,\bh) E^{\ah+\bh} & \mbox{for $\ah +\bh$ being a root}.
\end{cases} \nn
\eea
One sees immediately that the condition \eq{cond-cc1} is precisely the same 
condition  obtained from the Jacobi identity 
$[ [E^\ah, E^\bh], E^{-\ah}] + \cdots =0$ with $\ah+\bh \neq 0$; and 
the condition \eq{cond-cc4} 
is precisely the same condition obtained from the Jacobi identity 
$[E^\ah, [E^\bh, E^\gh]] + \cdots =0$ with
$\ah+\bh \neq 0$, $\bh+\gh \neq 0$, $\ah+\gh \neq 0$. 
Therefore the conditions \eq{cond-cc1} and \eq{cond-cc4} can be solved if 
$\ah$  and $c(\ah,\bh)$ are given by those
of a semisimple Lie algebra. 

We remark on passing that the conditions \eq{cond-cc1}, \eq{cond-cc4}  
do not apply if there is only a single pair of roots.

This concludes our
analysis of the consistency conditions for  Cartan-Weyl 3-algebras.

\hspace{2cm}
\\
\centerline{**** ****}

Summarizing, a  Cartan-Weyl 3-algebra is given by the relations 
\bea 
\; [H_I, H_J, H_K] &=& 0, \label{cw01} \\
\; [H_I, H_J, E^\a]& =& \a_{IJ} E^\a, \label{cw02}
\eea
\begin{subnumcases}{[H_I,E^\a, E^\b] =}
\a_{IK} g^{KL} H_L, & \mbox{if $\a+ \b =0$,} \label{cw03a}\\
g_I(\a,\b) E^{\a+\b}, & \mbox{if $\a+ \b \neq 0$ is a root,} \label{cw03b}\\
0,  & \mbox{if $\a+ \b$ is not a root,} \label{cw03c}
\end{subnumcases}
\begin{subnumcases}{[E^\a,E^\b, E^\g] =}
- g_K(\a,\b) g^{KL} H_L,  & \mbox{if $\a+ \b + \g =0$,} \label{cw04a} \\
0,  & \mbox{if $\a+ \b + \g \neq 0$ a root,} 
 \label{cw04b}\\
0,  & \mbox{if $\a+ \b + \g$ is not a root.}   \label{cw04c}
\end{subnumcases} 
If the algebra has only a single pair of roots
$\pm \a$, then \eq{cond1} is the only condition to be satisfied.
Otherwise, in general the root space can be decomposed into a number
of (say $M$) components: $\D_\cA(\cH) = \oplus_{\cci=1}^M \Omega_\cci$, 
where there is a null 
vector $\ph^{(\cci)}$ associated with each $\Omega_\cci$. The roots in each
$\Omega_\cci$ can be decomposed in the form
\be \label{cond-a}
\boxed{
\a = \ph^{(\cci)} \wedge \ah^{(\cci)}, \quad \a \in \Omega_\cci
}
\ee
and 
\be 
\boxed{
g_I (\a,\b)  = \begin{cases}
\ph_I^{(\cci)} c^{(\cci)}(\ah^{(\cci)},\bh^{(\cci)}), & \quad \mbox{for $\a,\b \in
\Omega_\cci$}, \\
0, & \quad \mbox{otherwise},
\end{cases}
}
\ee
where the one-form parts $\ah^{(\cci)}$ form the root system 
of a semisimple Lie algebra $\mg^{(\cci)}$.
The one-forms $\ph^{(\cci)}$ and $\ah^{(\ccj)}$ satisfy the conditions
\eq{cond6a}-\eq{cond7} and
the coefficient 
\if
$c(\a,\b)$ takes the form
\be \label{cond-c}
\boxed{
c(\a,\b) = \begin{cases}
c^{(\cci)}(\ah^{(\cci)},\bh^{(\cci)}),
& \mbox{$\a,\b \in \Omega_i$, }
 \\
0, & \mbox{otherwise},
\end{cases}
}
\ee 
where 
\fi
$c^{(\cci)}$ specifies the ``$[E,E]=E$''  type brackets of 
the semisimple Lie algebra $\mg^{(\cci)}$: 
\bea \label{lie-cw}
\; && [H_{\Ih_\cci},H_{\Jh_\ccj}] =0, \nn\\
\; && [H_{\Ih_\cci},E^{\ah^{(\cci)}}] = \ah_{\Ih_\cci} E^{\ah^{(\cci)}}, \\
\; && [E^{\ah^{(\cci)}},E^{\bh^{(\cci)}}] = \begin{cases}
\ah^{(\cci)}\cdot H & \mbox{for ${\ah^{(\cci)}}+ {\bh^{(\cci)}} =0$},\\
c^{(\cci)}(\ah^{(\cci)}, \bh^{(\cci)}) E^{\ah^{(\cci)}+ \bh^{(\cci)}} 
& \mbox{for ${\ah^{(\cci)}}+ {\bh^{(\cci)}}$  being a root}.
\end{cases} \nn
\eea

Therefore we have reduced the problem of classifying Cartan-Weyl
3-algebra to the problem of constructing the null vectors $\ph^{(\cci)}$ and 
roots $\ah^{(\cci)}$ such that \eq{cond6a}-\eq{cond7} are satisfied. This final
step will depend on the signature of the metric of the 3-algebra and will
be carried out in the main text of the paper.


\end{document}